\newif\ifonecol
\onecoltrue

\ifonecol
    
    \documentclass[12pt,draftclsnofoot,onecolumn]{IEEEtran}
\else
    \documentclass[journal,comsoc]{IEEEtran}
\fi

\usepackage{amsmath,graphics,amssymb,epsfig,subfigure,color,amsthm,cite}
\usepackage{array}
\usepackage{multirow}
\usepackage{enumerate}
\usepackage{algorithmic}
\usepackage{algorithm}
\usepackage{bm}
\usepackage[cmintegrals]{newtxmath}

\newtheorem{thm}{Theorem}

\newcommand{\argmax}{\operatornamewithlimits{argmax}}
\newcommand{\argmin}{\operatornamewithlimits{argmin}}
\makeatletter
\newcommand{\vast}{\bBigg@{3.5}}
\newcommand{\Vast}{\bBigg@{4.5}}
\makeatother

\begin{document}

\ifonecol
	\title{\LARGE{One-Bit Sphere Decoding for Uplink Massive MIMO Systems with One-Bit ADCs}}
\else
	\title{One-Bit Sphere Decoding for Uplink Massive MIMO Systems with One-Bit ADCs}
\fi

\author{Yo-Seb Jeon, Namyoon Lee, Song-Nam Hong, and Robert W. Heath, Jr.
\thanks{Y.-S. Jeon and N. Lee are with the Department of Electrical Engineering, POSTECH, Pohang, Gyeongbuk, 37673 Korea  (e-mail: yoseb.jeon@postech.ac.kr, nylee@postech.ac.kr).}
\thanks{S.-N. Hong is with the Department of Electrical and Computer Engineering, Ajou University, Suwon, Gyeonggi, 16499 Korea (e-mail: snhong@ajou.ac.kr).}
\thanks{R. W. Heath, Jr. is with the Wireless Networking and Communications Group, 	The University of Texas at Austin, Austin, TX 78712 USA (e-mail: rheath@utexas.edu).}}

\vspace{-2mm}

\maketitle

\begin{abstract} 
This paper presents a low-complexity near-maximum-likelihood-detection (near-MLD) algorithm called \textit{one-bit-sphere-decoding} for an uplink massive multiple-input multiple-output (MIMO) system with one-bit analog-to-digital converters (ADCs). The idea of the proposed algorithm is to estimate the transmitted symbol vector sent by uplink users (a codeword vector) by searching over a sphere, which contains a collection of codeword vectors close to the received signal vector at the base station in terms of \textit{a weighted Hamming distance}. To reduce the computational complexity for the construction of the sphere, the proposed algorithm divides the received signal vector into multiple sub-vectors each with reduced dimension. Then, it generates multiple spheres in parallel, where each sphere is centered at the sub-vector and contains a list of sub-codeword vectors. The detection performance of the proposed algorithm is also analyzed by characterizing the probability that the proposed algorithm performs worse than the MLD. The analysis shows how the dimension of each sphere and the size of the sub-codeword list are related to the performance-complexity tradeoff achieved by the proposed algorithm. Simulation results demonstrate that the proposed algorithm achieves near-MLD performance, while reducing the computational complexity compared to the existing MLD method.
\end{abstract}

\section{Introduction}
 Wireless systems with ultra low-precision analog-to-digital converter (ADC) are a power and cost efficient solution for future cellular networks that support wide bandwidths and a large number of antennas at the base station (BS) \cite{Murmann,Walden1999,Nossek2006,MezghaniNossek2007,Madhow2009,Mo2015,Liang2016,Jacobsson2017}. For the multiple-input multiple-output (MIMO) system with $B$-bit ADCs, where $B>1$, finding an optimal data detection method is challenging.  The challenge arises from multiple quantization levels at the ADCs, which can be differently chosen to minimize the detection error probability according to an input constellation and the variance of noise. Numerous sub-optimal data detection and channel estimation methods have been proposed assuming fixed quantization levels \cite{Dabeer2010,Wang2014,Wang2015,Studer2016,Hong2017,Jeon2017,Wen2016}.

 The use of one-bit ADCs in MIMO systems is interesting from both practical and theoretical perspectives \cite{Mo2015,Risi2016,Choi2016,Mollen2017,Li2017}. One major implementation advantage is the simplification of the circuit complexity by removing automatic gain control \cite{Singh2009}. In addition, the characterization of the channel capacity becomes tractable due to a fixed quantization level (e.g., zero-threshold comparator). For example, some capacity bounds of the MIMO system with one-bit ADCs were characterized when employing channel state information at the transmitter for a noise-free case \cite{Mo2015}.  Beside a capacity characterization, it is also possible to analytically derive the maximum likelihood detection (MLD) for the MIMO systems with one-bit ADCs \cite{Choi2016}, which yields the minimum error probability of detecting transmit symbols.

 The MLD problem for MIMO systems with one-bit ADCs differs from that for the MIMO system with infinite-precision ADCs. The MLD for the conventional MIMO systems under Gaussian noise reduces to the minimum Euclidean distance detection problem over a finite constellation set \cite{Damen2003,Kailath2006,Choi2010,Guo2006}. In contrast, the MLD for the MIMO system with one-bit ADCs finds an integer vector that maximizes the product of Q-functions \cite{Choi2016} instead of solving the least-squares problems. Nevertheless, the computational complexity of both MLD problems is NP-hard due to the integer constraint on the feasible set.

 Some low-complexity detection methods have been developed for MIMO systems with one-bit ADCs \cite{Risi2016,Choi2016,Mollen2017}. For instance, a heuristic zero-forcing detection (ZFD) method using one-bit quantized measurements was introduced in \cite{Risi2016}. A drawback of ZFD is that the number of receive antennas should be much larger than all possible numbers of transmit symbol vectors to reliably detect the transmitted data symbols. In other words, for a given number of receive antennas, the constellation size and/or the number of uplink users sending the data should be small to achieve a target level of the data detection performance. A near-MLD method using a convex relaxation technique was proposed in \cite{Wang2014}. This approach was also extended to devise ML data detection and channel estimation methods for the MIMO systems employing one-bit ADCs \cite{Choi2016}. The common idea is to convert the non-convex optimization problem that finds the optimal integer vector to a convex-optimization problem, and then find the solution using gradient-decent algorithms.

 Sphere decoding is a low-complexity detection method for MIMO systems with infinite-precision ADCs \cite{Damen2003,Kailath2006,Choi2010,Guo2006,FinckePohst1985,Hassibi2005}. The basic idea of sphere decoding is to search over only integer input vectors that lie in a certain sphere of radius $d$ around an initial estimate of input vector ${\bf \hat x}$. It diminishes the computational complexity by reducing the search space, while achieving near-MLD performance. The sphere decoding algorithms in \cite{Damen2003,Kailath2006,Choi2010,Guo2006}, do not extend to MIMO system with one-bit ADCs. This is because they find a set of integer vectors within the sphere with radius of $d$ in terms of the Euclidian distance between a received vector ${\bf y}$ and the product of the channel matrix ${\bf H}$ and the initial estimate of transmit vector ${\bf \hat x}$ \cite{Damen2003,Kailath2006,Choi2010}. When employing one-bit ADCs, the initial estimate of ${\bf \hat x}$ with one-bit measurements is inaccurate using ZFD. Furthermore, constructing the sphere using the Euclidian distance is not optimal when the received signal at the BS is quantized, as proven in \cite{Hong2017} for one-bit ADCs.

  In this paper, a low-complexity detection algorithm inspired by sphere decoding is presented for an uplink massive MIMO system with one-bit ADCs. The major contributions of this paper are summarized as follows.

  \begin{itemize}

   	\item We develop a near-optimal detection method for the uplink of multi-user MIMO system with one-bit ADCs. The proposed method is a variant of minimum weighted-Hamming-distance detection (MWD) that was originally introduced in \cite{Hong2017}. Unlike \cite{Hong2017} in which the weights are defined in an integral form, the proposed MWD exploits closed-form weights when computing weighted-Hamming-distances by approximating the Q-function.

   \item  We propose a low-complexity near-MLD algorithm called \textit{one-bit-sphere-decoding (OSD)}. The key idea of the OSD is to perform the proposed MWD over a sphere, which is a reduced set of all possible symbol (codeword) vectors that are close to the received signal at the BS in terms of the weighted Hamming distance. To diminish the computational complexity for the construction of codeword list in the sphere, we divide the received signal vector into multiple sub-vectors each with a reduced dimension. Then, we generate multiple spheres in parallel, where each sphere is centered at the sub-vector and contains a list of sub-codeword vectors.
   We compare the detection complexities between the proposed OSD and the existing MLD, and show the gains in the deduction of the complexity for OSD over MLD.

  	\item  We quantify the detection performance loss of the proposed OSD compared to the optimal performance. To this end, we characterize an upper bound of the probability that the proposed OSD performs worse than the MWD.
  	In the characterization, we first show that this probability is upper bounded by the sphere-list-error-probability (SEP), which is the probability that the index of the transmitted codeword does not belong to the constructed list in the sphere. We then derive an analytical expression for the upper bound of the SEP in terms of the relevant system parameters: 1) the number of uplink users, 2) the number of receive antennas at the BS, 3) the size of the codeword list in the sphere, and 4) the dimension of the sub-vector. Our result reveals how the multi-user detection error behaves with these relevant parameters.

  	\item Using simulations, we compare the detection performance of the OSD with those of the MLD and the MWD for both uncoded and coded MIMO systems with one-bit ADCs. Simulation results show that for the uncoded system, the OSD has near-MLD detection performance, while achieving a significant reduction in the detection complexity compared to the MLD. For the coded MIMO system, the OSD is implemented with a soft-output decoder by applying the technique in \cite{Kim-Hong2017} and is shown to achieve a significant frame-error-rate (FER) reduction compared to a hard-output decoder.
  \end{itemize}

\subsubsection*{Notation}
Upper-case and lower-case boldface letters denote matrices and column vectors, respectively.
$\mathbb{E}[\cdot]$ is the statistical expectation,
    $\mathbb{P}(\cdot)$ is the probability,
    $(\cdot)^\top$ is the transpose,
    $|\cdot|$ is the absolute value,
    $\text{Re}(\cdot)$ is the real part,
    $\text{Im}(\cdot)$ is the imaginary part,
    and $\lfloor\cdot\rfloor$ is the floor function.
${\bf 1}_n$ is an $n$-dimensional vector whose elements are all ones.
 ${\mathbb I}(A)$ is an indicator function that equals one if an event $A$ is true and zero otherwise.

\section{System Model}\label{sec:model}
In this section, we present a model for an uplink massive MIMO system with one-bit ADCs and provide definitions that will be used in the sequel.

\begin{figure}[t]
	\centering
	\epsfig{file=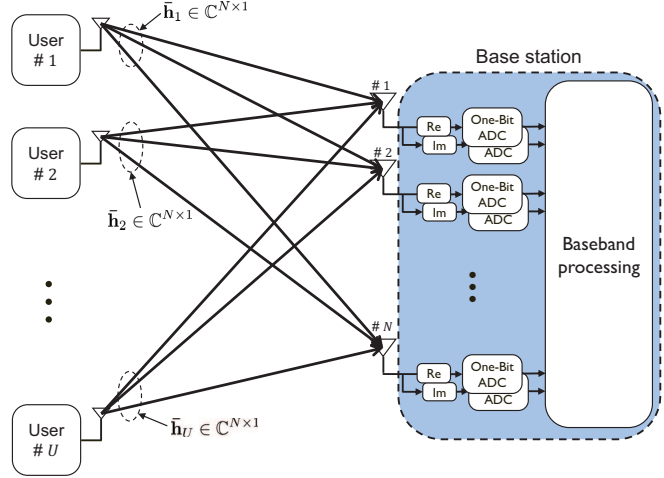, width=9cm}
	\caption{Illustration of a $U$-user uplink massive MIMO system that operates with one-bit ADCs.}
	\label{fig:System}
\end{figure}

\subsection{System Model}
We consider an uplink massive MIMO system in which $U$ uplink users, each equipped with a single transmit antenna, send data symbols to a BS equipped with $N$ receive antennas, as illustrated in Fig.~\ref{fig:System}. We denote a (data) symbol vector sent by the uplink users at time slot $t$ as ${\bf \bar x}[t] =[{\bar x}_1[t], {\bar x}_2[t],\cdots, {\bar x}_{U}[t]]^{\top}\in \mathbb{C}^{U}$, where each data symbol ${\bar x}_u[t]$ is drawn from a constellation set $\mathcal{\bar X}$ with size $M$, i.e., $|\mathcal{\bar X}|=M$, and satisfies $\mathbb{E}[|{\bar x}_u[t]|^2]=1$. In other words, the symbol vector ${\bf \bar x}[t]$ collects the transmitted signals from all users at time slot $t$. We define $\mathcal{X}$ as a constellation set for real or imaginary part such that $\mathcal{\bar X} = \{x_{\rm R}  + j x_{\rm I}~|~ x_{\rm R},x_{\rm I} \in \mathcal{X}\}$ and $|\mathcal{X}|=\sqrt{M}$. Then $\mathcal{X}^{2U}$ represents a symbol vector set that contains all possible combinations of transmit symbols sent by the $U$ uplink users.

We assume a frequency-flat MIMO channel. Let ${\bf \bar h}_{u} \in\mathbb{C}^{N\times 1}$ be the channel vector from the $u$th uplink user to the BS. Then, the channel impulse response is given by a channel matrix ${\bf \bar H}=[{\bf \bar h}_1,\ldots, {\bf \bar h}_{U}]\in\mathbb{C}^{N\times U}$. We also make two assumptions: 1) perfect synchronization at the BS and 2) perfect power control across all uplink users. Under these assumptions, the received signal vector at time slot $t$ before ADC quantization is
\begin{align}
    {\bf \bar r}[t] = {\bf \bar H}{\bf \bar x}[t] + {\bf \bar z}[t] \in \mathbb{C}^{N},  \label{eq:receive}
\end{align}
where ${\bf \bar z}[t]=[{\bar z}_1[t], {\bar z}_2[t],\cdots, {\bar z}_{N}[t]]^{\top}$ has elements independently drawn from a complex Gaussian distribution with zero mean and variance $\sigma^2$. We assume a block fading model in which the channel is time-invariant during coherence time interval. We denote $T_{\rm d}$ as the duration (the number of time slots) for data detection. The complex received signal in \eqref{eq:receive} can be equivalently rewritten in a real-form as
\begin{align}\underbrace{
\left[ {\begin{array}{c}
   {\rm Re}({\bf \bar r}[t])  \\ {\rm Im}({\bf \bar  r}[t]) \\
 \end{array} } \right]}_{{\bf r}[t]}=\underbrace{\left[ {\begin{array}{cc}
  {\rm Re}({\bf \bar H})& -{\rm Im}({\bf \bar H}) \\
   {\rm Im}({\bf \bar H}) & {\rm Re}({\bf \bar H})\\
 \end{array} } \right]}_{{\bf H}}\underbrace{\left[ {\begin{array}{c}
   {\rm Re}({\bf \bar x}[t])  \\ {\rm Im}({\bf \bar x}[t]) \\
 \end{array} } \right]}_{{\bf x}[t]}+\underbrace{\left[ {\begin{array}{c}
   {\rm Re}({\bf \bar z}[t])  \\ {\rm Im}({\bf \bar z}[t]) \\
 \end{array} } \right]}_{{\bf z}[t]},  \label{eq:receive2}
\end{align}
where ${\rm Re}({\bf A})$ and ${\rm Im}({\bf A})$ denote the real and complex parts of a complex matrix ${\bf A}$, respectively.

We consider the use of one-bit ADCs at each receive antenna, which implies that the real and the imaginary components of the received signal are separately quantized to binary levels. In this paper, ${\rm sign}(\cdot)$ is the quantization function, which essentially maps a positive value to 1 and a negative value to -1. Then the received signal after the ADCs at time slot $t$ is defined as ${\bf y}[t] = \left[y_1[t],y_2[t]\cdots, y_{2N}[t]\right]^\top \in \{+1,-1\}^{2N}$ with
\begin{align}
	{y}_{i}[t]={\rm sign}({r}_{i}[t]) = {\rm sign}\left({\bf h}_i^\top{\bf x}[t] + {z}_i[t]\right),  \label{eq:receive_real}
\end{align}
for all $i \in \mathcal{I} = \{1,2,\ldots,2N\}$, where ${\bf h}_i^\top$ is the $i$-th row of the channel matrix ${\bf H}$, and $z_i[t]$ is the $i$-th element of ${\bf z}[t]$.

\subsection{Definitions}
We provide some definitions that will be used in the sequel.

\vspace{1mm}
{\bf Definition 1 (Codewords and codebook \cite{Hong2017,Jeon2017})}: We define ${\bf c}_k = {\rm sign}({\bf H} {\bf x}_k)\in\{-1,1\}^{2N}$ as the $k$-th (binary) codeword vector corresponding to the $k$-th symbol vector ${\bf x}_k\in \mathcal{X}^{2U}$. For $k \in \mathcal{K}=\big\{1,2,\ldots,K=M^U\big\}$, each codeword vector ${\bf c}_k={\rm sign}({\bf H} {\bf x}_k)$ can be interpreted as a noise-free received signal when a symbol vector ${\bf x}_k \in \mathcal{X}^{2U}$ is transmitted via the channel matrix ${\bf H}$. We also define a codebook by a collection of codeword vectors, i.e., $\mathcal{C}=\{{\bf c}_1,{\bf c}_2,\ldots, {\bf c}_{K}\}$.  Note that similar notions for the codeword vectors and the codebook are considered in \cite{Hong2017,Jeon2017}.

\vspace{1mm}
{\bf Definition 2 (Weighted Hamming distance between codewords)}:
Let ${\bf c}_k=[c_{k,1},c_{k,2},\cdots,c_{k,2N}]^\top$ and ${\bf c}_j=[c_{j,1},c_{j,2},\cdots,c_{j,2N}]^\top$ be binary codeword vectors in a codebook $\mathcal{C}$.  In addition, let ${\bf w} = [w_1,w_2,\cdots, w_{2N} ]^\top\in \mathbb{R}^{2N}$ and ${\bf \tilde w}=[{\tilde w}_1,{\tilde w}_2,\cdots,{\tilde w}_{2N}]^\top\in \mathbb{R}^{2N}$ be weight vectors that consist of positive elements. The weight vector ${\bf w}$ is assigned when measuring the distance between the elements of ${\bf c}_k$ and ${\bf c}_j$ that have different signs. Whereas, the weight vector ${\bf \tilde w}$ is assigned when measuring the distance between the elements of ${\bf c}_k$ and ${\bf c}_j$ that have the same sign. Then, the weighted Hamming distance between ${\bf c}_k$ and ${\bf c}_j$ with respect to  ${\bf w}$ and  ${\bf \tilde w}$ is defined as
\begin{align}
    d_{\rm w}({\bf c}_k,{\bf c}_j;{\bf w},{\bf \tilde w}) \!=\! \sum_{i=1}^{2N} w_i \|c_{k,i}\!-\!c_{j,i}\|_0 \!+\!\! \sum_{i=1}^{2N} {\tilde w}_i (1-\|c_{k,i}\!-\!c_{j,i}\|_0),\label{eq:WD}
\end{align}
where $ \| {\bf a}\|_0$ is the zero norm that denotes the number of nonzero elements in a vector ${\bf a}$. Note that when ${\tilde w}_i > 0$ for any $i$, the weighted Hamming distance between two same codewords can be non-zero.

\section{MLD for MIMO System with One-Bit ADCs}\label{sec:MLDMWD}
In this section, we first review MLD for uplink massive MIMO systems with one-bit ADCs. We then show that the MLD is equivalent to MWD by leveraging the weighted Hamming distance defined in Section II-B. We finally develop a new MWD method which tightly approximates the MWD that is equivalent to the MLD. The developed MWD will be used as a baseline for a low-complexity near-MLD algorithm in Section~\ref{sec:main}.


\subsection{Maximum-Likelihood Detection (MLD)}
We present the MLD for uplink multi-user MIMO systems with one-bit ADCs that was originally introduced in \cite{Choi2016}. Let $p\left({\bf y}[t]|{\bf x}_k\right)$ be the likelihood function with the received signal ${\bf y}[t]$ when the $k$-th symbol vector, ${\bf x}_k \in \mathcal{X}^{2U}$, was sent at time slot $t$. Then, $p\left({\bf y}[t]|{\bf x}_k\right)$ is given by \cite{Choi2016}
\begin{align}
	 p\left({\bf y}[t]|{\bf x}_k\right)
	   &=\prod_{i=1}^{2N} p\left({y}_i[t]
    	|{\bf x}_k\right) =\prod_{i=1}^{2N} \left( 1- Q\left( \sqrt{\frac{2}{\sigma^2}}
		y_i[t]{\bf h}_i^\top {\bf x}_k  \right)\right),   \label{eq:likelihood}
\end{align}
where $Q(x)=\int_{x}^{\infty}\frac{1}{\sqrt{2\pi}} e^{-\frac{t^2}{2}}{\rm d}t$ is a standard Q-function.
Using \eqref{eq:likelihood}, MLD for a MIMO system with one-bit ADCs is represented by the following optimization problem:
\begin{align}
    {\bf \hat x}_{\rm MLD}[t]
    &= \underset{{\bf x}_k \in \mathcal{X}^{2U}}{\argmax}~
    	p\left({\bf y}[t]|{\bf x}_k\right)  \label{eq:defMLD} \\
    &= \underset{{\bf x}_k \in \mathcal{X}^{2U}}{\argmax}~
    	\prod_{i=1}^{2N} \left(1- Q\left( \sqrt{\frac{2}{\sigma^2}}
		y_i[t]{\bf h}_i^\top {\bf x}_k  \right) \right).   \label{eq:oneMLD}
\end{align}

\subsection{Minimum Weighted-Hamming-Distance Detection (MWD) as an Exact MLD}\label{sec:exactMWD}
We show that the MLD is equivalent to MWD. To this end, we demonstrate that the log-likelihood function can be rewritten in the form of the weighted Hamming distance in \eqref{eq:WD}. Let $\mathcal{N}_{k}^{\rm e}[t] = \{ i: y_i[t] \neq c_{k,i}\}$ be the index set of the received signal elements that have different signs with the elements of the $k$-th codeword vector under the premise that ${\bf x}_k \in \mathcal{X}^{2U}$ was sent at time slot $t$. Using this index set, we rewrite the likelihood function in \eqref{eq:likelihood} as
\begin{align}
{p}\left({\bf y}[t]|{\bf x}_k\right)
&= \prod_{i \in \mathcal{N}_{k}^{\rm e}[t]}
Q\left( - \sqrt{\frac{2}{\sigma^2}}y_i[t]{\bf h}_i^\top {\bf x}_k \right)
\prod_{i \notin \mathcal{N}_{k}^{\rm e}[t]}
\left( 1- Q\left( \sqrt{\frac{2}{\sigma^2}}
y_i[t]{\bf h}_i^\top {\bf x}_k  \right)\right),  \label{eq:ori_likelyhood}
\end{align}
where the equality is obtained by applying the property of the Q-function: $Q(x)=1-Q(-x)$. Then we take the logarithm of \eqref{eq:ori_likelyhood} which yields
	\begin{align}
	\ln\left(p\left({\bf y}[t]|{\bf x}_k\right) \right)
	&=\sum_{i \in \mathcal{N}_{k}^{\rm e}[t]}
	\ln Q\left( - \sqrt{\frac{2}{\sigma^2}}y_i[t]{\bf h}_i^\top {\bf x}_k \right)
	+ \sum_{i \notin \mathcal{N}_{k}^{\rm e}[t]}
	\ln\left(  1- Q\left( \sqrt{\frac{2}{\sigma^2}}
		y_i[t]{\bf h}_i^\top {\bf x}_k  \right) \right). \label{eq:ori_logLF}
	\end{align}
To simplify, we define two weights $w_{k,i}^\prime $ and ${\tilde w}_{k,i}^\prime $ as
\begin{align}
w_{k,i}^\prime \triangleq -\ln Q\left( - \sqrt{\frac{2}{\sigma^2}}y_i[t]{\bf h}_i^\top {\bf x}_k \right) > 0, \label{eq:ori_weight1}
\end{align}
and
\begin{align}
{\tilde w}_{k,i}^\prime \triangleq -\ln\left(  1- Q\left( \sqrt{\frac{2}{\sigma^2}} y_i[t]{\bf h}_i^\top {\bf x}_k  \right) \right) > 0. \label{eq:ori_weight2}
\end{align}
Using these weights, the log-likelihood function in \eqref{eq:ori_logLF} can be expressed as
\begin{align}
\ln\left(p\left({\bf y}[t]|{\bf x}_k\right) \right)
&=-\sum_{i \in \mathcal{N}_{k}^{\rm e}[t]}w_{k,i}^\prime  - \sum_{i \notin \mathcal{N}_{k}^{\rm e}[t]}  {\tilde w}_{k,i}^\prime\nonumber \\
&=-\sum_{i=1}^{2N}{w}_{k,i}^\prime
\left\|{y}_i[t] - c_{k,i}\right\|_0  -\sum_{i=1}^{2N}{\tilde w}_{k,i}^\prime
(1-\left\|{y}_i[t] + c_{k,i} \right\|_0), \label{eq:ori_logLF1}
\end{align}
where the last equality holds because
\begin{align*}
\left\|{y}_i[t] - c_{k,i}\right\|_0 = \begin{cases}
1,& i \in \mathcal{N}_{k}^{\rm e}[t],\\
0,& i \notin \mathcal{N}_{k}^{\rm e}[t].\\
\end{cases}
\end{align*}
From Definition 2 in Section II-B, the log-likelihood function is expressed in the form of the weighted Hamming distance:
\begin{align}
\ln\left(p\left({\bf y}[t]|{\bf x}_k\right) \right)
&= - d_{\rm w}\left({\bf y}[t],{\bf c}_k;{\bf w}_k^\prime,{\bf \tilde w}_k^\prime\right). \label{eq:ori_logLF2}
\end{align}
By applying the result in \eqref{eq:ori_logLF2} to the definition of MLD, we can show that the MWD is equivalent to the MLD:
\begin{align}
{\bf \hat x}_{\rm MLD}[t]
= \underset{{\bf x}_k \in \mathcal{X}^{2U}}{\argmax}~
\ln \left(p\left({\bf y}[t]|{\bf x}_k\right)\right)
=\underset{{\bf x}_k \in \mathcal{X}^{2U}}{\argmin}~
d_{\rm w}\left({\bf y}[t],{\bf c}_k;{\bf w}_k^\prime,{\bf \tilde w}_k^\prime\right).  \label{eq:oriMWD}
\end{align}

\subsection{MWD as a Near MLD}\label{sec:approxMWD}
Based on the MLD representation in \eqref{eq:oriMWD}, we develop a new MWD method that provides a near-MLD solution. The key idea of the developed MWD is to use closed-form weight vectors that tightly approximate the weight vectors in \eqref{eq:oriMWD}. In this approximation, we adopt a Q-function approximation in \cite{QBound} which demonstrates that ${\hat Q}(x)= \frac{1}{2}e^{-0.374x^2-0.777x}$ tightly approximates the Q-function for non-negative $x$ with the absolute error less than $10^{-3}$, i.e.,
\begin{align}
	|Q(x)-{\hat Q}(x)| \leq 10^{-3},~\text{for}~x\geq 0.   \label{eq:lem1:error1}
\end{align}
By applying $Q(x)\approx \frac{1}{2}e^{-0.374x^2-0.777x}$ for $x\geq 0$ to both \eqref{eq:ori_weight1} and \eqref{eq:ori_weight2}, we obtain two closed-form weights $w_{k,i}$ and $w_{k,i}^\prime$ that approximate the original weights ${w}_{k,i}^\prime$ and ${\tilde w}_{k,i}^\prime$, respectively, i.e.,
\begin{align}
  w_{k,i}^\prime &\approx w_{k,i}  \triangleq \frac{2a}{\sigma^2} |{\bf h}_i^{\top}{\bf x}_k|^2 +\frac{b\sqrt{2}}{\sigma} |{\bf h}_i^{\top}{\bf x}_k| + \ln 2>0,   \label{eq:weight1}   \\
  {\tilde w}_{k,i}^\prime &\approx {\tilde w}_{k,i}  \triangleq -\ln\left( 1-e^{-w_{k,i}}\right)>0, \label{eq:weight2}
\end{align}
where $a=0.374$ and $b=0.777$. By defining ${\bf w}_k = [w_{k,1},\cdots,w_{k,2N}]^\top$ and ${\bf \tilde w}_k = [{\tilde w}_{k,1},\cdots,{\tilde w}_{k,2N}]^\top$, we also obtain an approximation of the weighted Hamming distance in \eqref{eq:oriMWD} as
\begin{align}
d_{\rm w}\left({\bf y}[t],{\bf c}_k;{\bf w}_k^\prime,{\bf \tilde w}_k^\prime \right) \approx
d_{\rm w}\left({\bf y}[t],{\bf c}_k;{\bf w}_k,{\bf \tilde w}_k \right), \label{eq:WHD_approx}
\end{align}
for $k\in\mathcal{K}$. By leveraging the above approximation, we develop the MWD method that has closed-form weights, unlike the MWD in \eqref{eq:oriMWD}. The detection rule for the developed MWD method is given by
\begin{align}
	{\bf \hat x}_{\rm MWD}[t]  = \underset{{\bf x}_k \in \mathcal{X}^{2U}}{\argmin}~
	d_{\rm w}\left({\bf y}[t],{\bf c}_k;{\bf w}_{k},{\bf \tilde w}_{k}\right),  \label{eq:MWD}
\end{align}
where two weight vectors in \eqref{eq:MWD} can be computed at the BS from \eqref{eq:weight1} and \eqref{eq:weight2} when CSIR is available. The developed MWD in \eqref{eq:MWD} is expected to provide near-MLD solution because the developed MWD tightly approximates the MWD in~\eqref{eq:oriMWD} which has been shown to be equivalent to the MLD in the previous subsection. The main advantage of the developed MWD is that it does not require the Q-function calculation which necessarily relies on a mapping table for the computation in a practical system.

\vspace{1mm}
{\bf Remark 1 (Comparison to MLD for a conventional MIMO system):}
For a conventional MIMO system where the input-output relation is linear under Gaussian noise, the optimal MLD is equivalent to a minimum Euclidean distance detection. For a MIMO system with one-bit ADCs where the input-output relation is non-linear,  minimizing the weighted Hamming distance obtains near-MLD performance. Specifically, in the MIMO system with one-bit ADCs, the weighted Hamming distance is measured between noisy and quantized received signal, ${\bf y}[t]$, and noise-free but quantized received signal, ${\bf c}_k = {\rm sign}\left({\bf H}{\bf x}_k\right)$ for $k \in \mathcal{K}$. When measuring the distance between the $i$-th elements of ${\bf y}[t]$ and ${\bf c}_k$, different weights are assigned by taking into account both 1) the sign alignment between $y_i[t]$ and $c_{k,i}$ and 2) the reliability information provided by a received SNR $\frac{|{\bf h}_i^{\top}{\bf x}_k|^2}{\sigma^2}$. More precisely, if $y_i[t]\neq c_{k,i}$, the weight $w_{k,i}$ in \eqref{eq:weight1} is assigned which is an increasing function of the received SNR. Whereas, if $y_i[t]=c_{k,i}$, the weight ${\tilde w}_{k,i}$ in \eqref{eq:weight2} in assigned which is a decreasing function of the received SNR.

{\bf Remark 2 (High SNR regime):} When the received SNR is sufficiently large, ${\tilde w}_{k,i}$ approaches zero. This fact implies that when computing the weighted distance at high SNR, the receiver can ignore the elements of ${\bf y}[t]$ and ${\bf c}_k$ that have different signs. In this case, the weighted Hamming distance is computed as $ \sum_{i=1}^{2N}{w}_{k,i} \big\|{y}_i[t] - c_{k,i} \big\|_0$. This motivates us to further simplify the detection rule for MWD as follows:
\begin{align}\label{eq:MWD_high}
	{\bf \hat x}_{\rm MWD}[t]
	&\approx \underset{k \in \mathcal{K}}{\argmin}~  d_{\rm w}\left({\bf y}[t],{\bf c}_k;{\bf w}_k,{\bf 0}\right)
   = \underset{k \in \mathcal{K}}{\argmin}~ \sum_{i=1}^{2N} {w}_{k,i} \left\|  y_i[t]-c_{k,i}\right\|_0.
\end{align}
As can be seen in \eqref{eq:MWD_high}, the computational complexity of the MWD in \eqref{eq:MWD} can be reduced because the weights are computed only for the elements corresponding to $y_i[t] \neq {\rm sign}({\bf h}_i^\top {\bf x}_k)$.


\section{One-Bit Sphere Decoding}\label{sec:main}
In this section, based on the MWD in Section III-C, we propose a low-complexity near-MLD algorithm for uplink massive MIMO systems with one-bit ADCs, referred to as \textit{one-bit sphere decoding (OSD)}. We also compare the computational complexity of the proposed OSD with those of the MLD and the MWD, to show a significant reduction in the complexity achieved by the proposed OSD.


\subsection{Proposed Algorithm} \label{sec:OSD}
The key idea of the OSD is to construct a list of codeword vectors in the sphere for each possible received signal and then to perform the MWD only over the codeword list in the sphere. The major differences of the OSD to conventional sphere decoding algorithms in \cite{Damen2003,FinckePohst1985,Hassibi2005,Guo2006,Choi2010} are two folds:
\begin{itemize}
	\item The list of codeword vectors in the sphere is constructed using preprocessing based on CSIR. This preprocessing is only possible when the BS receives the signal vector in a finite set due to the use of one-bit ADCs.  This differs from the conventional sphere decoding algorithms in which the codeword list is constructed during data detection processing.
	\item The OSD measures the weighted Hamming distance when constructing the codeword list,  the conventional sphere decoding algorithms measure the Euclidean distance when finding the codewords in the sphere.
\end{itemize}


The OSD consists of two parts: \textit{list construction in the sphere} and \textit{detection over the sphere}. Detailed procedures of each part are given below.

\vspace{1mm}
{\bf  List construction in the sphere:}
The receiver constructs and saves a list of codeword vectors in the sphere for each received signal, using preprocessing based on CSIR. This list contains the indices of the codeword vectors that are close to the received signal in terms of the weighted Hamming distance. If the codeword list is constructed for all possible received signals, i.e., $\{{\bf y} \in \{-1,+1\}^{2N}\}$, the receiver requires to construct total $2^{2N}$ codeword lists. Then the complexity of the list construction could not be affordable in a practical system when $N$ is large. To resolve this problem, our strategy is to divide the received signal into $G\geq 1$ sub-vectors, each with dimension of $N_{\rm s} = \frac{2N}{G}$, and then to construct $2^{N_{\rm s}}$ possible sub-lists for each sub-vector ${\bf y} \in \{-1,+1\}^{N_{\rm s}}$ in parallel.


Let ${\bf y}_p^{(g)} \in \{-1,+1\}^{N_{\rm s}}$ be the $p$-th possible vector of the $g$-th sub-vector, where $p\in\{1,2,\ldots,2^{N_{\rm s}}\}$ and $g=\{1,2,\ldots, G\}$. Also let $\mathcal{I}_g$ be the index set for the elements of the $g$-th sub-vector, namely,
\begin{align}
	\mathcal{I}_g = \{(g-1)N_{\rm s}+1,(g-1)N_{\rm s}+2,\ldots,gN_{\rm s}\},~\text{for}~g \in \{1,2,\ldots,G\}.
\end{align}
Using this index set, the $g$-th sub-vector of ${\bf c}_k$ is defined as
\begin{align}
	{\bf c}_k^{(g)} = \big[c_{k,\mathcal{I}_g(1)},c_{k,\mathcal{I}_g(2)},\cdots,c_{k,\mathcal{I}_g(N_{\rm s})}\big]^\top,
\end{align}
while the weight vectors associated with ${\bf c}_k^{(g)}$ are defined as
\begin{align}
	{\bf w}_k^{(g)} &= \big[w_{k,\mathcal{I}_g(1)},w_{k,\mathcal{I}_g(2)},\cdots,w_{k,\mathcal{I}_g(N_{\rm s})}\big]^\top,~~\text{and}~~
	{\bf \tilde w}_k^{(g)} &= \big[{\tilde w}_{k,\mathcal{I}_g(1)},{\tilde w}_{k,\mathcal{I}_g(2)},\cdots,{\tilde w}_{k,\mathcal{I}_g(N_{\rm s})}\big]^\top, \label{eq:weight_g2}
\end{align}
respectively. Note that the above weight vectors can be computed at the receiver from \eqref{eq:weight1} and \eqref{eq:weight2} when CSIR is available. Let $\pi_g(\ell,p) \in \mathcal{K}$ be an index function indicating that ${\bf c}_{\pi_g(\ell,p)}^{(g)}$ is the $\ell$-th closest sub-codeword vector to ${\bf y}_p^{(g)}$, i.e.,
\ifonecol
\begin{align}
d_{\rm w}\!\left({\bf y}_p^{(g)},      {\bf c}_{\pi_g(\ell,p)}^{(g)}; {\bf w}_{\pi_g(\ell,p)}^{(g)}, {\bf \tilde w}_{\pi_g(\ell,p)}^{(g)} \right) \leq d_{\rm w}\!\left({\bf y}_p^{(g)}, {\bf c}_{\pi_g(t,p)}^{(g)};    {\bf w}_{\pi_g(t,p)}^{(g)},    {\bf \tilde w}_{\pi_g(t,p)}^{(g)}\right), \nonumber
\end{align}
\else
\begin{align}
	&d_{\rm w}\!\left({\bf y}_p^{(g)},      {\bf c}_{\pi_g(\ell,p)}^{(g)}; {\bf w}_{\pi_g(\ell,p)}^{(g)}, {\bf \tilde w}_{\pi_g(\ell,p)}^{(g)} \right) \nonumber \\
	&\leq d_{\rm w}\!\left({\bf y}_p^{(g)}, {\bf c}_{\pi_g(t,p)}^{(g)};    {\bf w}_{\pi_g(t,p)}^{(g)},    {\bf \tilde w}_{\pi_g(t,p)}^{(g)}\right), \nonumber
\end{align}
\fi
for $t \in \{\ell+1,\ell+2,\ldots,K\}$. Then the sub-list associated with ${\bf y}_p^{(g)}$ is determined as the indices of the sub-codeword vectors that are the $L$ closest to ${\bf y}_p^{(g)}$, that is
\begin{align}
   	\mathcal{S}_g\left({\bf y}_p^{(g)},L
   	\right) = \left\{ {\pi_g(1,p)}, {\pi_g(2,p)}, \ldots, {\pi_g(L,p)} \right\}, \label{eq:sub_sphere}
\end{align}
where $p\in\left\{1,2,\ldots,2^{N_{\rm s}}\right\}$ and $g \in \{1,2,\ldots,G\}$. The overall list-construction procedure of the OSD is illustrated in Fig.~\ref{fig:Concept}. Note that this procedure is performed only once during a channel coherence block.

\vspace{1mm}
{\bf Detection over the sphere:}
During data detection processing, the receiver estimates the codeword vector (the transmitted symbol vector) by searching over the list in the sphere generated during the list-construction process. Specifically, the receiver performs the MWD in \eqref{eq:MWD} over the list, to find the codeword vector that has the minimum weighted Hamming distance to the received signal.

When the received signal, ${\bf y}[t]$, is observed at time slot $t$, the receiver divides ${\bf y}[t]$ into $G$ sub-vectors, namely $\left\{{\bf y}^{(1)}[t],{\bf y}^{(2)}[t],\ldots,{\bf y}^{(G)}[t]\right\}$. Then for each sub-vector ${\bf y}^{(g)}[t]$, the receiver obtains the sub-list of the $L$ nearest sub-codeword vectors, i.e., $\mathcal{S}_g\left({\bf y}^{(g)}[t],L\right)$, that is generated during the list construction process. Using the obtained $G$ sub-lists, the receiver generates a \textit{total} list of the codeword vectors as the union of these $G$ sub-lists, i.e.,
\begin{align}
	\mathcal{S}({\bf y}[t]) = \bigcup_{g=1}^G \mathcal{S}_g\left({\bf y}^{(g)}[t],L\right). \label{eq:sphere_set}
\end{align}
Note that the cardinality of $\mathcal{S}({\bf y}[t])$ is bounded between $L$ and $GL$, i.e., $L\leq |\mathcal{S}({\bf y}[t])| \leq GL$ where $G$ and $L$ are chosen to be $GL \ll K$.
Using $\mathcal{S}({\bf y}[t])$, the receiver finds the index of the transmitted symbol vector by applying the detection rule for the MWD in \eqref{eq:MWD} over $\mathcal{S}({\bf y}[t])$, i.e.,
\begin{align}
	{k_{\rm OSD}^\star}[t] &= \underset{ k \in \mathcal{S} ({\bf y}[t])  }{\argmin} d_{\rm w}\left({\bf c}_k,{\bf y}[t]; {\bf w}_{k},{\bf \tilde w}_{k}\right).\label{eq:OSD}
\end{align}
Once the best index is found, the receiver obtains the estimate of the transmitted symbol vector ${\bf \hat x}_{\rm OSD}[t] = {\bf x}_{k_{\rm OSD}^{\star}[t]}$. The overall detection procedure of the OSD is depicted in Fig.~\ref{fig:Concept}.

\ifonecol
\begin{figure*}
	\centering \vspace{-0.3cm}
	\includegraphics[width=16cm]{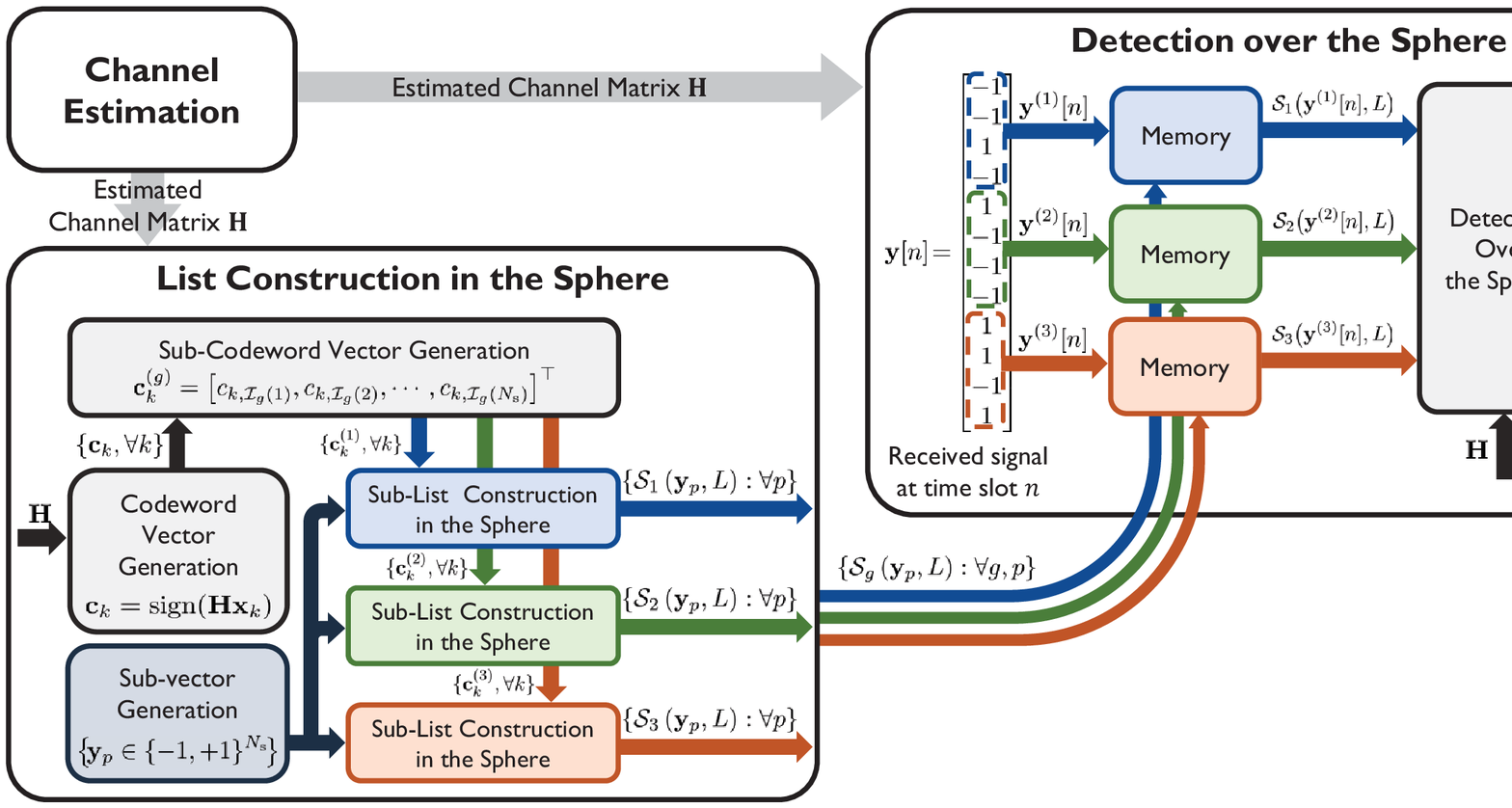} \vspace{-0.1cm}\caption{The proposed OSD consists of two parts: 1) list construction in the sphere and 2) detection over the sphere, when $N=6$, $N_{\rm s}=4$, and $G=3$.} \label{fig:Concept}
\end{figure*}
\else
\begin{figure*}
	\centering \vspace{-0.3cm}
	\includegraphics[width=18cm]{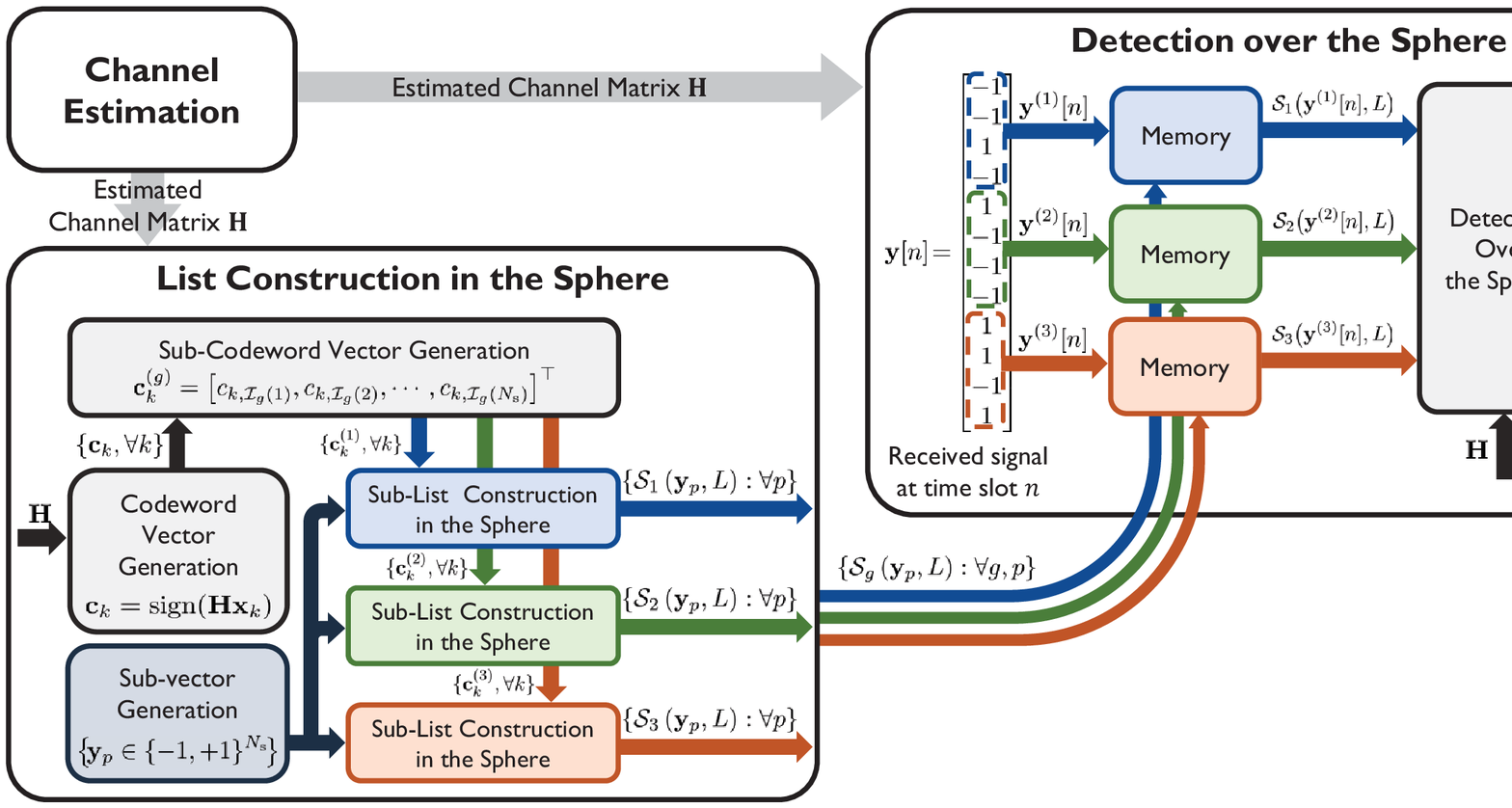} \vspace{-0.1cm}\caption{The proposed OSD consists of two parts: 1) list construction in the sphere and 2) detection over the sphere, when $N=6$, $N_{\rm s}=4$, and $G=3$.} \label{fig:Concept}
\end{figure*}
\fi

\vspace{3mm}
We present a simple example to illustrate the operation of the OSD.

{\bf Example 1:} Suppose a case in which $U=2$, $N=2$, $N_{\rm s}=2$, $L=1$, and BPSK modulation per user is assumed. We also consider a channel matrix that is given by
\begin{align}
{\bf H} = \left[\!\begin{array}{cc}
0.8 & 0.2 \\ 0.1 & 0.9 \\ -0.7 & 0.3 \\ 0.4 & -0.6 \\
\end{array}\!\right]
= \left[~\begin{array}{c}
{\bf h}_1^{\top}\\{\bf h}_2^{\top}\\{\bf h}_3^{\top}\\{\bf h}_4^{\top}
\end{array}~\right].
\end{align}
The receiver first constructs the list in the sphere. In this example, all possible transmit symbol vectors sent by the two uplink users are
\begin{align}
{\bf x}_1 =  \left[\!\begin{array}{c}
1\\1
\end{array}\!\right]\!,~
{\bf x}_2 = \left[\!\!\begin{array}{c}
1\\-1
\end{array}\!\!\right]\!,~
{\bf x}_{3} =  \left[\!\!\begin{array}{c}
-1\\1
\end{array}\!\!\right]\!,~
{\bf x}_{4} = \left[\!\begin{array}{c}
-1\\-1
\end{array}\!\right].
\end{align}
Since ${\bf c}_k ={\rm sign}({\bf H}{\bf x}_k)$ and $N_{\rm s}=2$,  the receiver generates four sub-codeword vectors as follows:
\begin{align}
{\bf c}_1^{(1)} = \left[\!\begin{array}{c}
1\\1
\end{array}\!\right]\!,~
{\bf c}_2^{(1)} = \left[\!\begin{array}{c}
1\\-1
\end{array}\!\right]\!,~
{\bf c}_3^{(1)} = \left[\!\begin{array}{c}
-1\\1
\end{array}\!\right]\!,~
{\bf c}_4^{(1)} = \left[\!\begin{array}{c}
-1\\-1
\end{array}\!\right],
\end{align}
and
\begin{align}
{\bf c}_1^{(2)} =\left[\!\begin{array}{c}
-1\\-1
\end{array}\!\right]\!,~
{\bf c}_2^{(2)} =\left[\!\begin{array}{c}
-1\\1
\end{array}\!\right]\!,~
{\bf c}_{3}^{(2)} =\left[\!\begin{array}{c}
1\\-1
\end{array}\!\right]\!,~
{\bf c}_{4}^{(2)} = \left[\!\begin{array}{c}
1\\1
\end{array}\!\right].
\end{align}
The receiver also generates four possible sub-vectors for the received signal:
\begin{align}
{\bf y}_1^{(g)} = \left[\!\begin{array}{c}
1\\1
\end{array}\!\right]\!,~
{\bf y}_2^{(g)} = \left[\!\begin{array}{c}
1\\-1
\end{array}\!\right]\!,~
{\bf y}_3^{(g)} = \left[\!\begin{array}{c}
-1\\1
\end{array}\!\right]\!,~
{\bf y}_4^{(g)} = \left[\!\begin{array}{c}
-1\\-1
\end{array}\!\right],   \label{eq:ex:y}
\end{align}
where ${\bf y}_p^{(g)}$ is the $p$-th possible vector in a finite set $\{-1,+1\}^{N_{\rm s}}$ and $g=\{1,2\}$. Because we consider $L=1$ case, the sub-list for ${\bf y}_p^{(g)}$ contains only one index of the sub-codeword vector ${\bf c}_{\pi(1,p)}^{(g)}$ that has the minimum weighted Hamming distance from ${\bf y}_p^{(g)}$. Suppose that eight sub-lists are
\begin{align}
& \mathcal{S}_1\left({\bf y}_1^{(1)},1\right) = \left\{ 1 \right \},  \mathcal{S}_1\left({\bf y}_2^{(1)},1\right) = \left\{ 2 \right \},
\mathcal{S}_1\left({\bf y}_3^{(1)},1\right) = \left\{ 3 \right \}, \mathcal{S}_1\left({\bf y}_4^{(1)},1\right) = \left\{ 4 \right \}, \nonumber \\
& \mathcal{S}_2\left({\bf y}_1^{(2)},1\right) = \left\{ 4 \right \}, \mathcal{S}_2\left({\bf y}_2^{(2)},1\right) = \left\{ 3 \right \},
\mathcal{S}_2\left({\bf y}_3^{(2)},1\right) = \left\{ 2 \right \}, \mathcal{S}_2\left({\bf y}_4^{(2)},1\right) = \left\{ 1 \right \}. \nonumber
\end{align}
The receiver constructs the codeword list in the sphere once for a channel coherence block.

Now, the receiver performs the detection over the codeword list. Suppose that at time slot $t$, the receiver observes a signal vector given by
${\bf y}[t] = [1,-1,-1,1]^\top$.
Then two sub-vectors corresponding to ${\bf y}[t]$ are ${\bf y}^{(1)}[t] = [1,-1]^\top$ and ${\bf y}^{(2)}[t] = [-1,1]^\top$.
Using these two sub-vectors, the receiver determines the codeword list in the sphere for ${\bf y}[t]$ as the union of $\mathcal{S}_1\left({\bf y}^{(1)}[t],1\right)$ and $\mathcal{S}_2\left({\bf y}^{(2)}[t],1\right)$, i.e.,
\begin{align}
\mathcal{S}({\bf y}[t]) = \bigcup_{g=1}^G \mathcal{S}_g\left({\bf y}^{(g)}[t],1\right)  = \left\{ 2,3 \right \}.
\end{align}
Then the receiver finds the index of the transmitted symbol vector by applying the detection rule for the MWD in \eqref{eq:MWD} over the codeword list in $\mathcal{S}({\bf y}[t])$:
\begin{align}
{k_{\rm OSD}^\star}[t] &= \underset{k \in \{2,3\}}{\argmin} ~d_{\rm w}\left({\bf c}_k,{\bf y}[t]; {\bf w}_{k},{\bf \tilde w}_{k}\right).
\end{align}
Finally, the receiver obtains the estimate of the transmitted symbol vector as ${\bf \hat x}_{\rm OSD}[t] = {\bf x}_{k_{\rm OSD}^{\star}[t]}$.

In this example, under the premise that each ${\bf y}[t]\in \{-1,+1\}^{2N}$ is generated with the equal probability, the average number of codeword search for the proposed OSD is
\begin{align}
   \frac{1}{2^{2N}} \sum_{{\bf y}[t]\in \{-1,+1\}^{2N}}|\mathcal{S}({\bf y}[t])| =1.75.
\end{align}
Therefore, the OSD achieves a 56$\%$ reduction of the computational complexity compared to the MWD which computes four different weighted Hamming distances for every ${\bf y}[t]\in \{-1,+1\}^{2N}$.

\vspace{1mm}
{\bf Remark 3 (The interplay between $N_{\rm s}$ and $L$):}
The dimension of sub-vector $N_{\rm s}$ and the list size $L$ determine the tradeoff between the detection performance and the computational complexity of OSD. If we set $N_{\rm s}$ to be large, the size of the sub-list, $L$, can be reduced because the weighted Hamming distance between ${\bf y}_p^{(g)}$ and ${\bf c}_{\pi(\ell,p)}^{(g)}$, i.e., $d_{\rm w}\!\!\left({\bf y}_p^{(g)},\! {\bf c}_{\pi(\ell,p)}^{(g)};\!{\bf w}_{\pi(\ell,p)}^{(g)} ,{\bf \tilde w}_{\pi(\ell,p)}^{(g)}\! \right)$ for large $N_{\rm s}$ provides enough information to reliably find the best codeword in the set $\mathcal{S}({\bf y}_p^{(g)}, L)$ with a small number of $L$. Whereas, if we set $N_{\rm s}$ to be small, the weighted Hamming distance between ${\bf y}_p^{(g)}$ and ${\bf c}_{\pi(\ell,p)}^{(g)}$ does not provide reliable information to correctly find the best codeword in the set $\mathcal{S}({\bf y}_p^{(g)}, L)$. Therefore, in this case, we need to choose a large size of $L$ to improve the detection performance. Note that one can also modify the algorithm by choosing a different dimension of $N_{\rm s}$ per sub-vector to further optimize the tradeoff between the detection performance and the computational complexity of the OSD.

{\bf Remark 4 (Extension to multi-precision ADCs):}
The proposed OSD can be extended for the case with multi-precision ADCs. Suppose that a $B$-bit scalar quantizer is independently applied to the real and imaginary parts of the received signal, while $\mathcal{Y} = \{q_1,q_2,\ldots,q_{2^B}\}$ is the set of all possible outputs of the quantizer, and ${\rm SQ}: \mathbb{R} \rightarrow \mathcal{Y}$ is the quantization function of the scalar quantizer.
In \cite{Studer2016}, it is shown that the log-likelihood function of this system is given by
\begin{align}
	\ln\left(p\left({\bf y}[t]|{\bf x}_k\right) \right)
	= \sum_{i=1}^{2N} \ln \left( Q\!\left(\frac{l(y_{i}[t]) - {\bf h}_i^\top {\bf x}_k}{\sigma/2}\right)
	 - Q\left(\frac{u({y}_{i}[t]) - {\bf h}_i^\top {\bf x}_k}{\sigma/2}\!\right) \right), \label{eq:multiLL}
\end{align}
where $u(y)$ and $l(y)$ are the upper and the lower bin boundaries associated with the quantized output $y\in\mathcal{Y}$. Define  ${\bf c}_k = {\rm SQ}({\bf H}{\bf x}_k)$ as the $k$-th codeword vector associating with the $k$-th symbol vector. Then, similar to \eqref{eq:ori_logLF}, the log-likelihood function in \eqref{eq:multiLL} is expressed as
\begin{align}
\ln\left(p\left({\bf y}[t]|{\bf x}_k\right) \right)
&= \sum_{i\notin \mathcal{N}_{k}^{\rm e}[t] } \ln \left( Q\!\left(\frac{l(c_{k,i}) - {\bf h}_i^\top {\bf x}_k}{\sigma/2}\right)
- Q\left(\frac{u(c_{k,i}) - {\bf h}_i^\top {\bf x}_k}{\sigma/2}\!\right) \right)  \nonumber \\
&~~~~~+\sum_{i\in \mathcal{N}_{k}^{\rm e}[t] } \ln \left( Q\!\left(\frac{l(y_{i}[t]) - {\bf h}_i^\top {\bf x}_k}{\sigma/2}\right)
- Q\left(\frac{u({y}_{i}[t]) - {\bf h}_i^\top {\bf x}_k}{\sigma/2}\!\right) \right) \nonumber \\
&\leq -\sum_{i\notin \mathcal{N}_{k}^{\rm e}[t] } \tilde{w}_{k,i}^\prime -\sum_{i\in \mathcal{N}_{k}^{\rm e}[t] } w_{k,i}^\prime
\nonumber \\
&= -d_{\rm w}\left({\bf y}[t],{\bf c}_k;{\bf w}_k^\prime,{\bf \tilde w}_k^\prime\right),\label{eq:multiLL3}
\end{align}
where $\mathcal{N}_{k}^{\rm e}[t] = \{ i: y_i[t] \neq c_{k,i}\}$, ${\bf w}_k^\prime = [w_{k,1},w_{k,2},\cdots, w_{k,2N} ]^\top$ with
\begin{align}
w_{k,i}^\prime &= - \max_{y \in \mathcal{Y}\!,~y\neq c_{k,i}} \ln \left( Q\!\left(\frac{l(y) - {\bf h}_i^\top {\bf x}_k}{\sigma/2}\right)
- Q\left(\frac{u({y}) - {\bf h}_i^\top {\bf x}_k}{\sigma/2}\!\right) \right),  \label{eq:multiweight1}
\end{align}
and  ${\bf \tilde w}_k^\prime=[{\tilde w}_{k,1},{\tilde w}_{k,2},\cdots,{\tilde w}_{k,2N}]^\top$ with
\begin{align}
\tilde{w}_{k,i}^\prime &= - \ln \left( Q\!\left(\frac{l(c_{k,i}) - {\bf h}_i^\top {\bf x}_k}{\sigma/2}\right)
- Q\left(\frac{u(c_{k,i}) - {\bf h}_i^\top {\bf x}_k}{\sigma/2}\!\right) \right).  \label{eq:multiweight2}
\end{align}
Motivated by the inequality in \eqref{eq:multiLL3}, the MLD of the MIMO systems with multi-precision ADCs can be approximated as
\begin{align}
{\bf \hat x}_{\rm MLD}[t]
= \underset{{\bf x}_k \in \mathcal{X}^{2U}}{\argmax}~
\ln \left(p\left({\bf y}[t]|{\bf x}_k\right)\right)
\approx \underset{{\bf x}_k \in \mathcal{X}^{2U}}{\argmin}~
d_{\rm w}\left({\bf y}[t],{\bf c}_k;{\bf w}_k^\prime,{\bf \tilde w}_k^\prime\right).  \label{eq:multiMWD}
\end{align}
Except for the definition of the weight vectors, the detection rule in \eqref{eq:multiMWD} is exactly the same with the detection rule in \eqref{eq:oriMWD}. Therefore, the proposed OSD can also be extended to the MIMO systems with multi-precision ADCs, simply by using the weights in \eqref{eq:multiweight1} and \eqref{eq:multiweight2} when computing the weighted Hamming distance. Note that although the detection rule in \eqref{eq:multiMWD} is an approximate MLD, it achieves the exact MLD as the number of precision bits at the ADCs decreases.

{\bf Remark 5 (Extension to frequency-selective channels):}
	Although the proposed OSD is developed under the assumption of frequency-flat channels, it can also be applied to frequency-selective channels with some modifications. Suppose that the number of channel-impulse-response (CIR) taps of the channel is given by $L\geq 1$. For this channel, consider multiple block transmissions; each consists of $B$ successive data symbols followed by $L-1$ zeros at the end. Then the received signal vector at the $t$-th time slot of the $b$-th block transmission is expressed as
	\begin{align}
	{\bf y}_b[t]&={\rm sign}\left(\sum_{\ell=0}^{L-1}{\bf H}[\ell]{\bf x}_b[t-\ell] + {\bf z}_b[t]\right), \label{eq:receive_real3}
	\end{align}
	where ${\bf H}[\ell] \in \mathbb{R}^{2N\times 2U}$ is the real-domain channel matrix that consists of the $\ell$-th CIR taps, and ${\bf x}_b[t] \in \mathbb{R}^{2U}$ and ${\bf z}_b[t] \in \mathbb{R}^{2N}$ are the transmitted symbol and the noise at the $t$-th time slot of the $b$-th block transmission, respectively. By concatenating the received signals during $T_{\rm b}+L-1$ time slots, the total received signal vector of the $b$-th block transmission is given by
	\begin{align}\label{eq:toepy}
		\left[ \begin{array}{c}
		{\bf y}[1] \\ {\bf y}[2] \\ \vdots \\ {\bf y}[B+L-2]  \\ {\bf y}[B+L-1]
		\end{array}\right]
		&= {\rm sign}\left(\left[\begin{array}{ccccc}
		{\bf H}[0] & ~~{\bf 0}     & ~~\cdots    & {\bf 0}       & {\bf 0} \\
		{\bf H}[1] & ~~{\bf H}[0]  & ~~\ddots    & ~             & \vdots \\
		\vdots     & ~~\ddots      & ~~\ddots    & ~             & ~  \\
		{\bf 0}    & ~           & ~         & {\bf H}[L-1]  & {\bf H}[L-2]   \\
		{\bf 0}    & ~~\cdots    & ~~\cdots    & {\bf 0}       & {\bf H}[L-1]  \\
		\end{array}\right]
		\left[\begin{array}{c}
		{\bf x}[1] \\ {\bf x}[2] \\  \vdots \\ {\bf x}[B-1] \\ {\bf x}[B] \\
		\end{array}\right]
		+
		\left[\begin{array}{c}
		{\bf z}[1] \\ {\bf z}[2] \\  \vdots \\ {\bf z}[B-1] \\ {\bf z}[B] \\
		\end{array}\right]
		\right).
		\end{align}		
	The received signal in \eqref{eq:toepy} is equivalent to the received signal of an uplink MIMO system that has $UB$ uplink users and $N(B+L-1)$ receive antennas at the BS. Therefore, the proposed OSD is directly applicable to frequency-selective channels by assuming that there exists $UL$ uplink users and $N(B+L-1)$ receive antennas. Note that the above extension requires a significant detection complexity when both $U$ and $B$ are large, even for the proposed OSD; thereby, as future work, it would be interesting to develop a lower complexity method for the use in a practical system with frequency-selective channels.

\begin{table*}
	\renewcommand{\arraystretch}{1.7}
	\caption{The number of real multiplications required for various detection methods when $N\gg1$.}
	\label{table:complexity}
	\centering
	\begin{tabular}{c||c|c}
		\hline
		\bfseries Detection method & Preprocessing
		& Data detection processing \\
		\hline\hline
		MLD in \eqref{eq:oneMLD}
		& -
		& $(4U + 6)NKT_{\rm d}$
		\\
		\hline
		MWD in \eqref{eq:MWD}
		& -
		& $(4U + 14)NKT_{\rm d}$
		\\	
		\hline
		OSD in \eqref{eq:OSD}
		& $2^{N_{\rm s}}(4U + 14)NK$
		& $\frac{2NL}{N_{\rm s}}(4U + 14)NT_{\rm d} $  \\	
		\hline
	\end{tabular}
\end{table*}

\subsection{Computational Complexity Comparison}\label{sec:complexity}
We compare the computational complexity of three detection methods: MLD, MWD, and OSD. To this end, we compute the number of real multiplications required for each method, which is summarized in Table~\ref{table:complexity}. Specifically, for the OSD, we consider the worst case in which the size of the codeword list in the sphere is maximized, i.e., $GL = \frac{2N}{N_{\rm s}}L$, as can be seen from~\eqref{eq:sphere_set}. Table~\ref{table:complexity} shows that the number of real multiplications required for the proposed OSD is $\Big(\frac{2NL}{N_{\rm s}K} + \frac{2^{N_{\rm s}}}{T_{\rm d}}\Big) \times \left(\frac{4U +14}{4U +6}\right)$ of that for the MLD and $\Big(\frac{2NL}{N_{\rm s}K} + \frac{2^{N_{\rm s}}}{T_{\rm d}}\Big)$ of that for the MWD, even in the worst case. These results imply that if two design parameters of the OSD, $L$ and $N_{\rm s}$, are properly set, the OSD has a less detection complexity than both the MLD and the MWD do. Therefore, by setting $L \ll K$ and ${N_{\rm s}} \ll \log_2 T_{\rm d}$, the proposed OSD achieves a significant reduction in the detection complexity compared to both methods. Note that a similar result also holds for the comparison of the numbers of real additions.

\section{Detection Performance of One-Bit Sphere Decoding}\label{sec:analysis}
In this section, we analyze the detection performance of the proposed OSD by characterizing an upper bound of the probability that the proposed OSD performs worse than the MWD. 
We first demonstrate that this probability is upper bounded by the sphere-list-error-probability (SEP), which is the probability that the index of the transmitted codeword does not belong to the constructed list in the sphere.

Let $P_{\rm loss}$ be the probability that the detection error occurs using the proposed OSD while the detection is correct using the MWD.  Then $P_{\rm loss}$ is expressed as
\ifonecol
	\begin{align}\label{eq:OSDequi}
	P_{\rm loss} &= \sum_{k=1}^{K} {\rm Pr}\left( {\bf \hat x}_{\rm MWD}[t] = {\bf x}_k,  {\bf \hat x}_{\rm OSD}[t]\neq {\bf x}_k,  {\bf x}[t] = {\bf x}_k \right) \nonumber \\
	&= \sum_{k=1}^{K}{\rm Pr}\left({\bf \hat x}_{\rm MWD}[t] = {\bf x}_k,  {\bf \hat x}_{\rm OSD}[t]\neq {\bf x}_k, k \in \mathcal{S}({\bf y}[t]), {\bf x}[t] = {\bf x}_k \right)  \nonumber \\
	&~~~+ \sum_{k=1}^{K} {\rm Pr}\left({\bf \hat x}_{\rm MWD}[t] = {\bf x}_k,  {\bf \hat x}_{\rm OSD}[t]\neq {\bf x}_k, k \notin \mathcal{S}({\bf y}[t]), {\bf x}[t] = {\bf x}_k \right).
	\end{align}
\else

\vspace{-3mm}{\small{\begin{align}\label{eq:OSDequi}
	&P_{\rm loss} = \sum_{k=1}^{K} {\rm Pr}\left( {\bf \hat x}_{\rm MWD}[t] = {\bf x}_k,  {\bf \hat x}_{\rm OSD}[t]\neq {\bf x}_k,  {\bf x}[t] = {\bf x}_k \right) \nonumber \\
	&= \sum_{k=1}^{K}{\rm Pr}\left({\bf \hat x}_{\rm MWD}[t] = {\bf x}_k,  {\bf \hat x}_{\rm OSD}[t]\neq {\bf x}_k, k \in \mathcal{S}({\bf y}[t]), {\bf x}[t] = {\bf x}_k \right)  \nonumber \\
	&~~~+ \sum_{k=1}^{K} {\rm Pr}\left({\bf \hat x}_{\rm MWD}[t] = {\bf x}_k,  {\bf \hat x}_{\rm OSD}[t]\neq {\bf x}_k, k \notin \mathcal{S}({\bf y}[t]), {\bf x}[t] = {\bf x}_k \right).
\end{align}}}\noindent
\fi
 By the detection rule for the OSD in \eqref{eq:OSD}, if the transmitted codeword index does not belong to the codeword list inside of the sphere, the OSD fails to detect the correct symbol vector; thereby, the error event $\{k \notin \mathcal{S}({\bf y}[t]), {\bf x}[t] = {\bf x}_k\}$ is a subset of the event $\{{\bf \hat x}_{\rm OSD}[t]\neq {\bf x}_k, {\bf x}[t] = {\bf x}_k\}$. In addition, if the MWD finds the transmitted symbol index, the OSD also finds the transmitted symbol index, provided that this index is in the codeword list in the sphere; thereby, the intersection of two event sets  ${\bf \hat x}_{\rm MWD}[t] = {\bf x}_k\}$ and $ \{ {\bf \hat x}_{\rm OSD}[t]\neq {\bf x}_k, k \in \mathcal{S}({\bf y}[t]) \}$ is an empty set. From this fact, we rewrite $P_{\rm loss}$ in \eqref{eq:OSDequi} as
\ifonecol
\begin{align}\label{eq:OSDequi2}
P_{\rm loss} &=  \sum_{k=1}^{K} {\rm Pr}\left({\bf \hat x}_{\rm MWD}[t] = {\bf x}_k,  k \notin \mathcal{S}({\bf y}[t]), {\bf x}[t] = {\bf x}_k \right) \nonumber \\
&\leq \sum_{k=1}^{K} {\rm Pr}\left(k \notin \mathcal{S}({\bf y}[t]), {\bf x}[t] = {\bf x}_k \right) =P_{\rm SEP},
\end{align}
\else

\vspace{-3mm}{\small{\begin{align}\label{eq:OSDequi2}
P_{\rm loss} &=  \sum_{k=1}^{K} {\rm Pr}\left({\bf \hat x}_{\rm MWD}[t] = {\bf x}_k,  k \notin \mathcal{S}({\bf y}[t]), {\bf x}[t] = {\bf x}_k \right) \nonumber \\
&\leq \sum_{k=1}^{K} {\rm Pr}\left(k \notin \mathcal{S}({\bf y}[t]), {\bf x}[t] = {\bf x}_k \right) =P_{\rm SEP},
\end{align}}}\noindent
\fi
where $P_{\rm SEP}$ is the SEP.

Now, we characterize the upper bound of the SEP using the following theorem.
\begin{thm}\label{Thm:SEP}
	For a fixed channel matrix ${\bf H} \in \mathbb{R}^{2N\times 2U}$, SEP of the proposed OSD is
	\begin{align}\label{eq:Thm2}
	P_{\rm SEP}
	&= \sum_{k=1}^{K} {\rm Pr}\left(k \notin \mathcal{S}\left( {\bf y}[t] \right), {\bf x}[t]={\bf x}_k\right)  \nonumber \\
	&\lessapprox \frac{1}{K}\sum_{k=1}^{K} \prod_{g=1}^{\frac{2N}{N_{\rm s}}} \sum_{{\bf e}\in \mathcal{E}_{k}^{(g)}(L)}
	\!\exp\left( - {\bf e}^\top {\bf w}_k^{(g)} - ({\bf 1}_{N_{\rm s}}-{\bf e})^\top {\bf \tilde w}_k^{(g)} \right),
	\end{align}
	where
	\begin{align}\label{eq:Thm2:E_kg}
	\mathcal{E}_{k}^{(g)}(L) = \left\{{\bf e} : d_{{\rm min},k}^{(g)}({\bf e},L) \leq {\bf 1}_{N_{\rm s}}^\top {\bf \tilde w}_{k}^{(g)}, {\bf e} \in \{0,1\}^{N_{\rm s}} \right\},
	\end{align}
	$d_{{\rm min},k}^{(g)}({\bf e},L)$ is the $L$-th smallest element of $\Big\{d_{k,j}^{(g)} + {\bf e}^\top {\bm \Delta}_{k,j}^{(g)}\Big\}_{j \neq k}$,
	$d_{k,j}^{(g)}= {d}_{\rm w}\left({\bf c}_{j}^{(g)},{\bf c}_k^{(g)};{\bf w}_{j}^{(g)}, {\bf \tilde w}_{j}^{(g)}\right)$,
	and ${\bm \Delta}_{k,j}^{(g)} \in \mathbb{R}^{N_{\rm s}}$ is a vector whose $i$-th element is ${\Delta}_{k,j,i}^{(g)} =  \left({w}_{j,i}^{(g)} - {\tilde w}_{j,i}^{(g)}\right)c_{k,i}c_{j,i}  -  \left({w}_{k,i}^{(g)} - {\tilde w}_{k,i}^{(g)}\right)$.
\end{thm}
\begin{IEEEproof}
	See Appendix~\ref{sec:Apdx1}
\end{IEEEproof}
\vspace{1mm}

\begin{figure}
	\centering
	\includegraphics[width=3.1in]{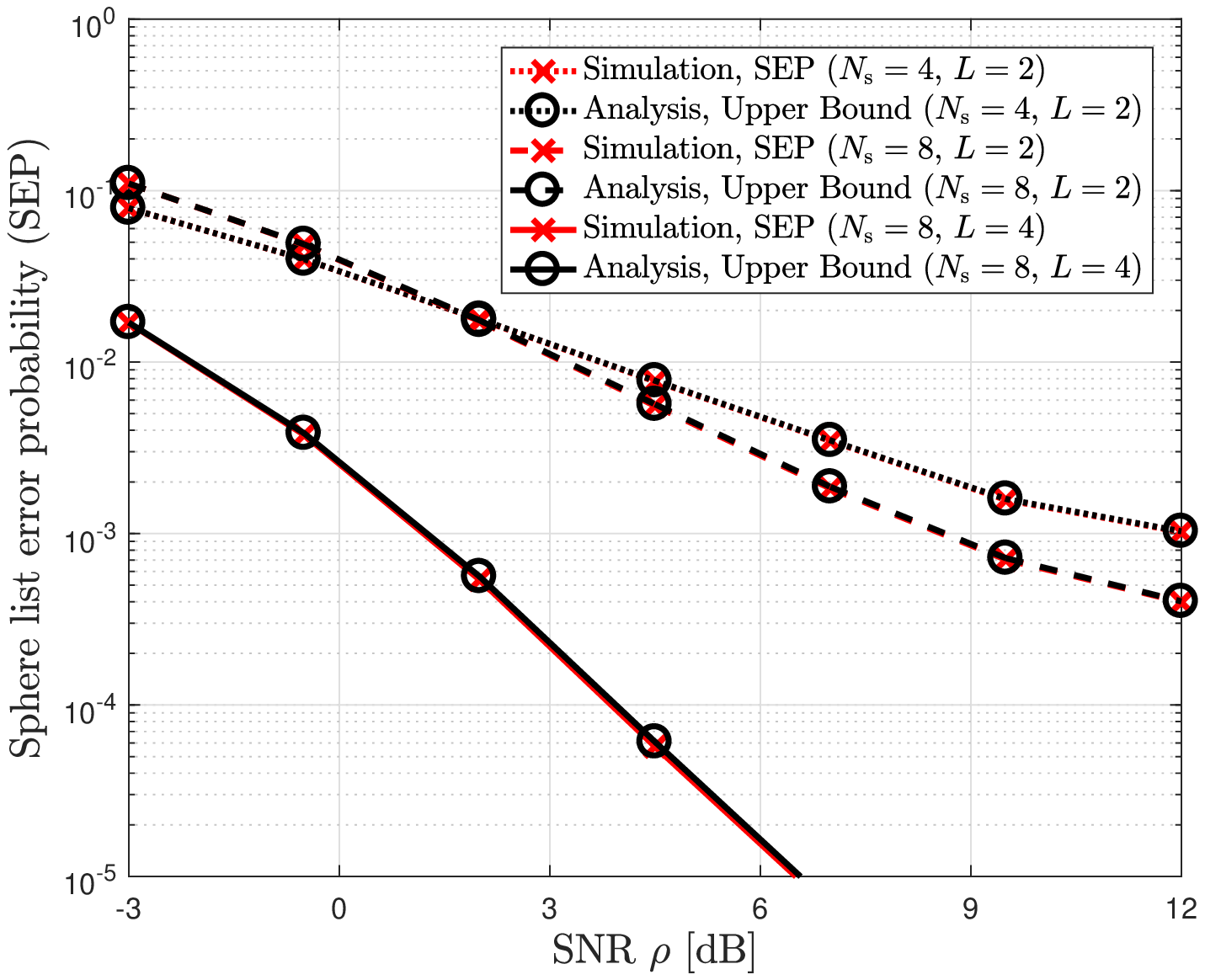} \vspace{-0.3cm}\caption{Comparison of the SEP obtained by simulations with the approximate upper bound calculated from \eqref{eq:Thm2} when $U=2$, $N=8$, 4-QAM is adopted, and CSIR is perfect.} \vspace{-3mm} \label{fig:SEP}
\end{figure}

From Theorem~\ref{Thm:SEP}, we can show that the SEP of the proposed OSD decreases with both the dimension of a sub-vector, $N_{\rm s}$, and the size of a sub-list in the sphere, $L$. To see this, it should first be noticed that the upper bound in \eqref{eq:Thm2} decreases with $d_{{\rm min},k}^{(g)}({\bf e},L)$ because the size of a set $\mathcal{E}_{k}^{(g)}(L)$ is reduced by increasing $d_{{\rm min},k}^{(g)}({\bf e},L)$. This parameter can be shown to be an increasing function of both $N_{\rm s}$ and $L$; first, $d_{{\rm min},k}^{(g)}({\bf e},L)$ is defined as the $L$-th smallest value, so $d_{{\rm min},k}^{(g)}({\bf e},L)$ increases with $L$; next, increasing the dimension of each sub-codeword vector increases the term $d_{k,j}^{(g)}$ in the definition of $d_{{\rm min},k}^{(g)}({\bf e},L)$, so $d_{{\rm min},k}^{(g)}({\bf e},L)$ also increases with $N_{\rm s}$.

We also present a numerical example to show the tightness of the approximate upper bound derived in Theorem~\ref{Thm:SEP}.


\vspace{1mm}
{\bf Example 2:} In Fig.~\ref{fig:SEP}, we compare the SEP obtained by simulations with the approximate upper bound of the SEP calculated by \eqref{eq:Thm2} when $U=2$, $N=8$, and 4-QAM is used. Simulation results are averaged over 5000 random realizations of channel coefficients that are independently drawn from a complex Gaussian distribution with zero mean and  unit variance. Channel stat information at the receiver (CSIR) is assumed to be perfect. Fig.~\ref{fig:SEP} shows that the approximate upper bound is very tight to the simulated SEP regardless of the values of $N_{\rm s}$ and $L$; thereby, this result validates our analysis in Theorem~\ref{Thm:SEP}. Another important observation in Fig.~\ref{fig:SEP} is that the SEP of the proposed OSD decreases with both $N_{\rm s}$ and $L$, as we have expected from Theorem 1. Specifically, it is shown that the SEP obtained when $(Ns; L) = (8;4)$ is significantly lower than the SEP obtained when $(Ns; L) = (4;2)$, while the computational complexity of the former case is only $12.5\%$ higher than that of the latter case when $T_{\rm d}=4096$ (see Table~\ref{table:complexity}). This observation implies that the determination of $N_{\rm s}$ and $L$ has a considerable impact on the tradeoff between the SEP and the computational complexity when using the proposed OSD.

\section{Simulation Results}\label{sec:simul}
In this section, using simulations, we evaluate the detection performance of the proposed OSD for uplink massive MIMO systems with one-bit ADCs. All simulation results are averaged over 5000 random realizations of channel coefficients that are independently drawn from a complex Gaussian distribution with zero mean and unit variance.

\subsection{Uncoded Performance}
We evaluate the detection performance of the proposed OSD for an uncoded system. For a comparison, we also present the performances of the conventional MLD in \eqref{eq:oneMLD} and the proposed MWD in \eqref{eq:MWD}.


\begin{figure}
    \centering
	\includegraphics[width=3.1in]{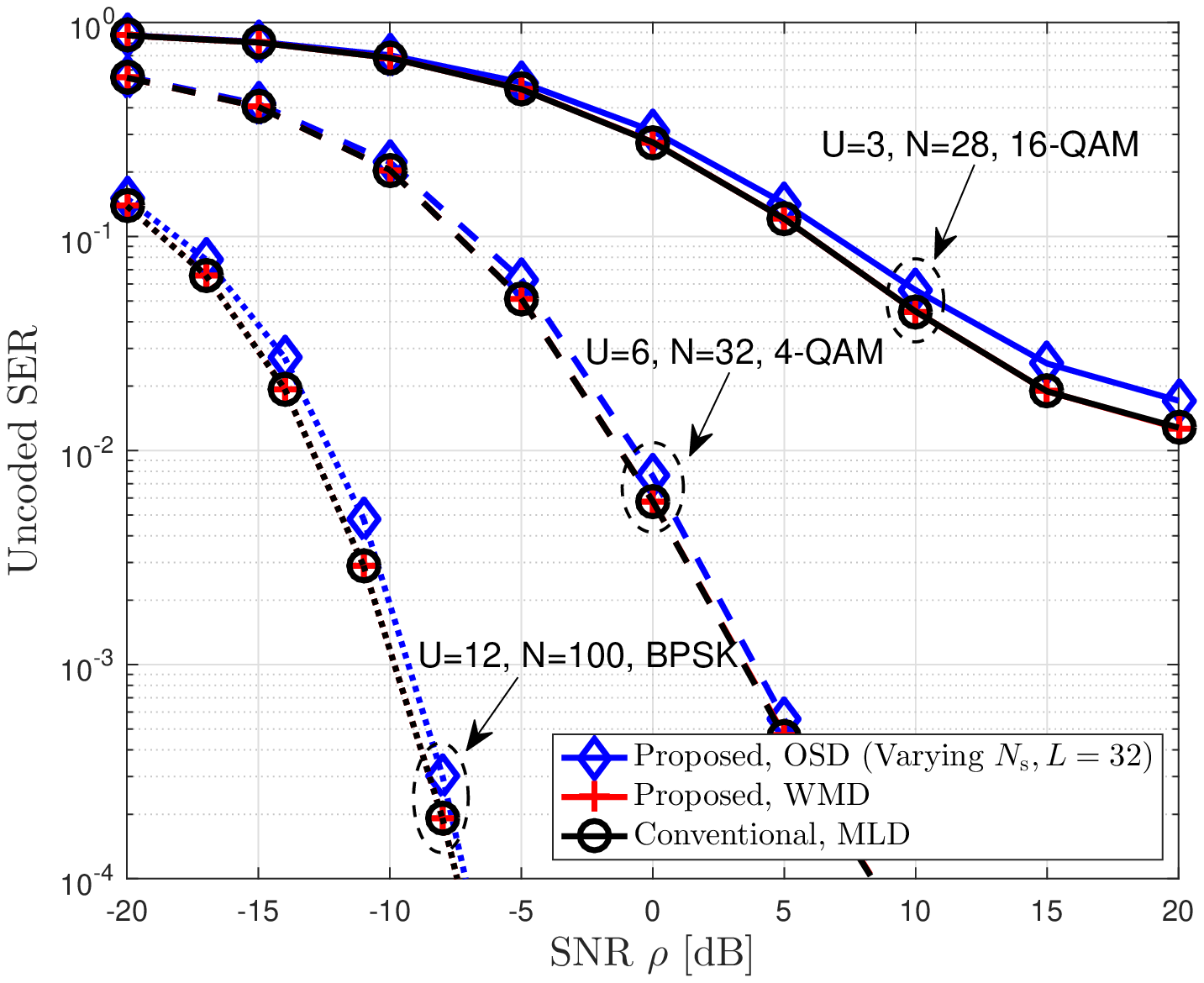} \vspace{-0.3cm}\caption{The SER vs. SNR of the proposed OSD, the proposed MWD, and the conventional MLD for various $U$, $N$, and constellation sets with perfect CSIR.} \vspace{-3mm} \label{fig:Nr}
\end{figure}

Fig.~\ref{fig:Nr} compares the symbol-error-rate (SER) of the OSD with those of the MLD and the MWD when CSIR is perfect. For the OSD, the dimension of the sub-codeword vector is set as $N_{\rm s}= 7, 8, 10$ for $N=28, 32, 100$, respectively. Fig.~\ref{fig:Nr} shows that the OSD has a negligible SER loss compared to the MLD regardless of the number of receive antennas, the number of users, the constellation set, and the SNR. Meanwhile,  the OSD reduces the detection complexity (i.e., the number of real multiplications) of the MLD by $89\%, 88\%, 68\%$ when $N=28, 32, 100$, respectively (for $T_{\rm d}=8192$ case, see Table~\ref{table:complexity}). 
It is also  noticeable that this complexity reduction further increases as $T_{\rm d}$ increases. These results show that the proposed OSD provides a good performance-complexity tradeoff for the uplink massive MIMO systems with one-bit ADCs. Although the MWD shows almost the same SER performance to the MLD, it does not provide any reduction in the detection complexity as seen in Table~\ref{table:complexity}.


\begin{figure}
    \centering
	\includegraphics[width=3.1in]{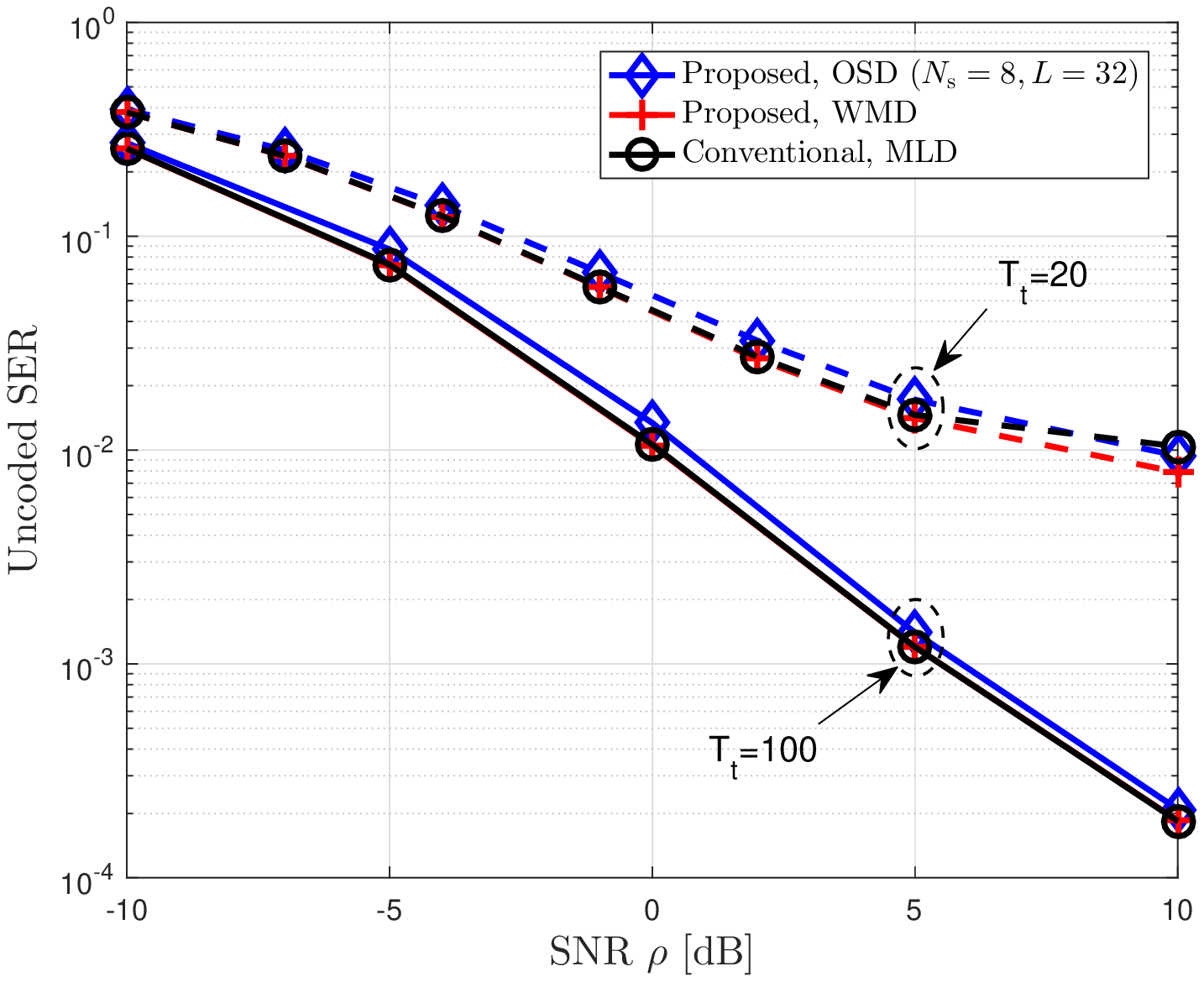} \vspace{-0.3cm}\caption{The SER vs. SNR of the proposed OSD, the proposed MWD, and the conventional MLD when pilot-based channel estimation is applied with various lengths of pilot signals for the case of $U=6$, $N=32$, and 4-QAM modulation.} \vspace{-3mm} \label{fig:CE}
\end{figure}

Fig.~\ref{fig:CE} compares the SER of the OSD with those of the MLD and the MWD when pilot-based channel estimation is applied with various lengths of pilot signals (i.e., $T_{\rm t}= 20$ or $100$). For the channel estimation, we apply a ML-based estimation method developed in \cite{Choi2016}. Fig.~\ref{fig:CE} shows that the OSD achieves a near-optimal SER with a reduced detection complexity, regardless of the length of pilot signals. This result implies that the improvement of the performance-complexity tradeoff achieved by the OSD is also robust to the channel estimation error.


\begin{figure}
	\centering
	\includegraphics[width=3.1in]{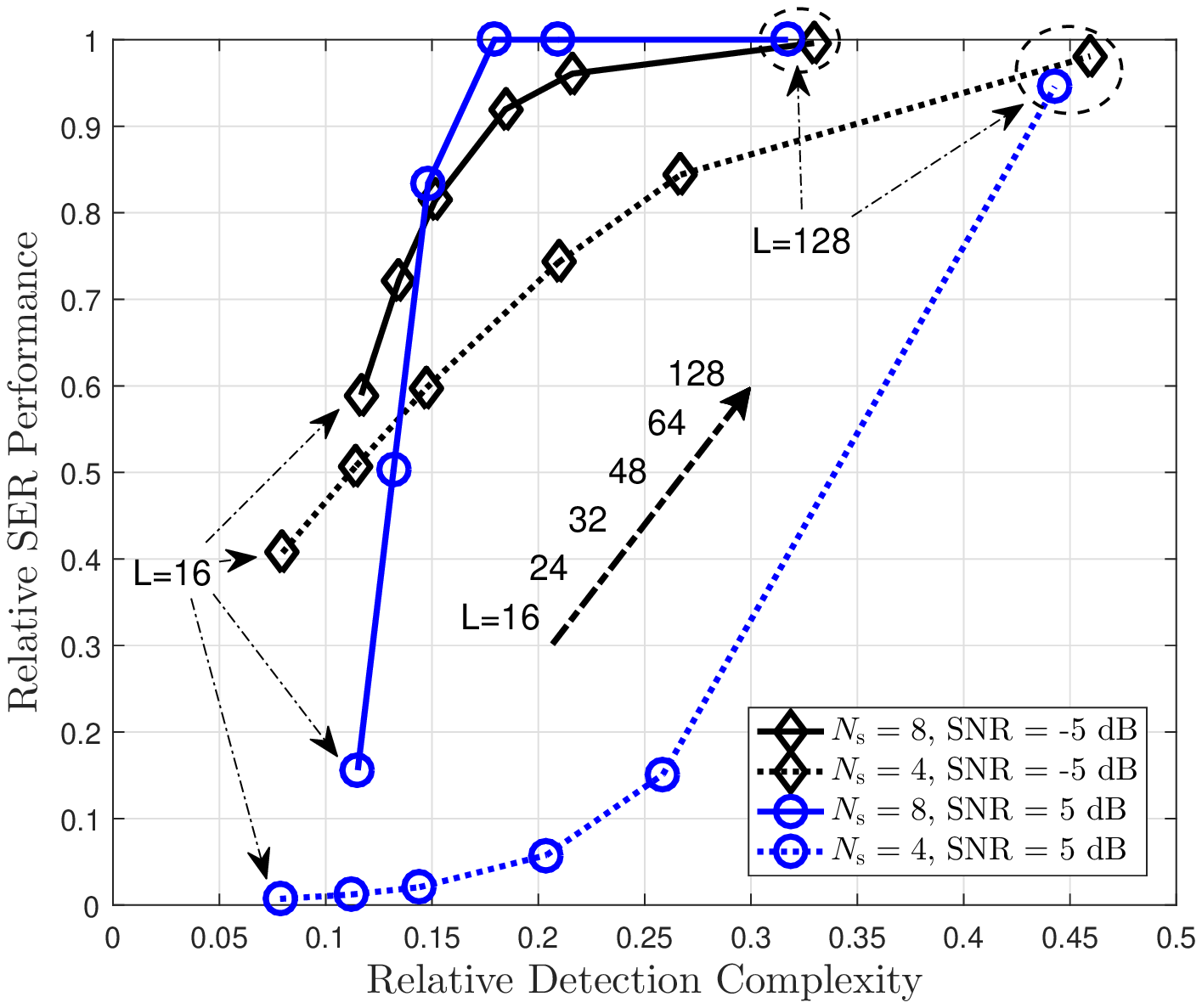} \vspace{-0.3cm}\caption{The performance-complexity tradeoff achieved by the proposed OSD for various $N_{\rm s}$ and $L$ with $U=6$, $N=32$, 4-QAM modulation, and perfect CSIR.} \label{fig:tradeoff}
\end{figure}


Fig.~\ref{fig:tradeoff} plots the performance-complexity tradeoff achieved by the OSD for various dimensions of a sub-codeword vector, $N_{\rm s}$,  and also for various sizes of a sub-list in the sphere, $L$, when CSIR is perfect. The relative SER performance in the $y$-axis is computed as the ratio of the SER achieved by the MLD to that achieved by the OSD, while a relative complexity in the $x$-axis is computed as the ratio of the number of real multiplications required by the OSD to that required by the MLD. Fig.~\ref{fig:tradeoff} shows that as $L$ increases, both the relative SER performance and the relative complexity increase; this result shows the tradeoff relation between the performance and the complexity when using the OSD. In addition, the detection performance of the OSD with $N_{\rm s}=8$ is much higher than that with $N_{\rm s}=4$, which implies that the performance-complexity tradeoff is very sensitive to the choice of $N_{\rm s}$. Another interesting observation is that the complexity required to achieve the optimal SER performance reduces as the SNR increases. Based on this observation, the effectiveness of the OSD can be improved with the operating SNR of the system.


\subsection{Coded Performance}
We also evaluate the detection performance of the proposed OSD for a coded system. As an underlying channel code, we adopt a 1/2-rate LDPC code of the blocklength $N_{\rm B}=672$ from the IEEE 802.11ad standardization \cite{IEEE}. For the conventional MLD, we only employ a {\em hard-input} bit-flipping decoder \cite{Rao} as in \cite{Choi2016}, since they produce hard-decision outputs. For both the MWD and the OSD, besides the hard-decision outputs, we also derive soft outputs using the technique in \cite{Kim-Hong2017}, which enables to use a {\em soft-input} belief-propagation decoder \cite{Richardson}. For completeness, we briefly explain how to compute soft outputs from the hard-decision measurement ${\bf y}[t]$ for the proposed OSD. Without loss of generality, we only focus on the $u$-th channel decoder to decode the user $u$'s message. Recall that  $\mathcal{S}({\bf y}[t])$ contains the all codewords in the search-space. We first partition the  $\mathcal{S}({\bf y}[t])$ into the four subsets which are defined as
\begin{equation}
\mathcal{S}^{u}({\bf y}[t]|i)=\{k \in \mathcal{S}({\bf y}[t]): {\bar x}_{k,u} = \bar{\mathcal{X}}(i)\},
\end{equation}
for $i\in\{1,2,3,4\}$,  where ${\bar x}_{k,u}$ denotes the $u$-th element of the ${\bf \bar x}_{k}$ and $\bar{\mathcal{X}}(i)$ denotes the $i$-th element of the 4-QAM constellation set $\bar{\mathcal{X}}$. Using this definition, the log-likelihood ratios (LLRs), i.e., the inputs of the belief-propagation decoder, are computed as
\begin{align*}
L_{2n-1}^{u}({\bf y}[t]) &= \min_{k \in \mathcal{S}^{u}({\bf y}[t]|2)\cup \mathcal{S}^{u}({\bf y}[t]|3)} d_{\rm w}({\bf y}[t], {\bf c}_{k};{\bf w}_{k}, \tilde{{\bf w}}_{k})\\
&~~~-\min_{k \in \mathcal{S}^{u}({\bf y}[t]|0)\cup \mathcal{S}^{u}({\bf y}[t]|1)} d_{\rm w}({\bf y}[t],{\bf c}_{k};{\bf w}_{k}, \tilde{{\bf w}}_{k})\\
L_{2n}^{u}({\bf y}[t]) &=\min_{k \in \mathcal{S}^{u}({\bf y}[t]|1)\cup \mathcal{S}^{u}({\bf y}[t]|3)} d_{\rm w}({\bf y}[t], {\bf c}_{k};{\bf w}_{k}, \tilde{{\bf w}}_{k})\\
&~~~-\min_{k \in \mathcal{S}^{u}({\bf y}[t]|0)\cup \mathcal{S}^{u}({\bf y}[t]|2)} d_{\rm w}({\bf y}[t], {\bf c}_{k};{\bf w}_{k}, \tilde{{\bf w}}_{k}),
\end{align*}
for $n\in \{1,\ldots,N_{\rm B}\}$. The resulting $2N_{\rm B}$ LLRs $\{L_{2n-1}^{u}({\bf y}[t]),L_{2n}^{u}({\bf y}[t]): n=1,\ldots,N_{\rm B}\}$ are embedded into the {\em soft} belief-propagation decoder as the soft inputs. 

\begin{figure}
	\centering
	\includegraphics[width=3.1in]{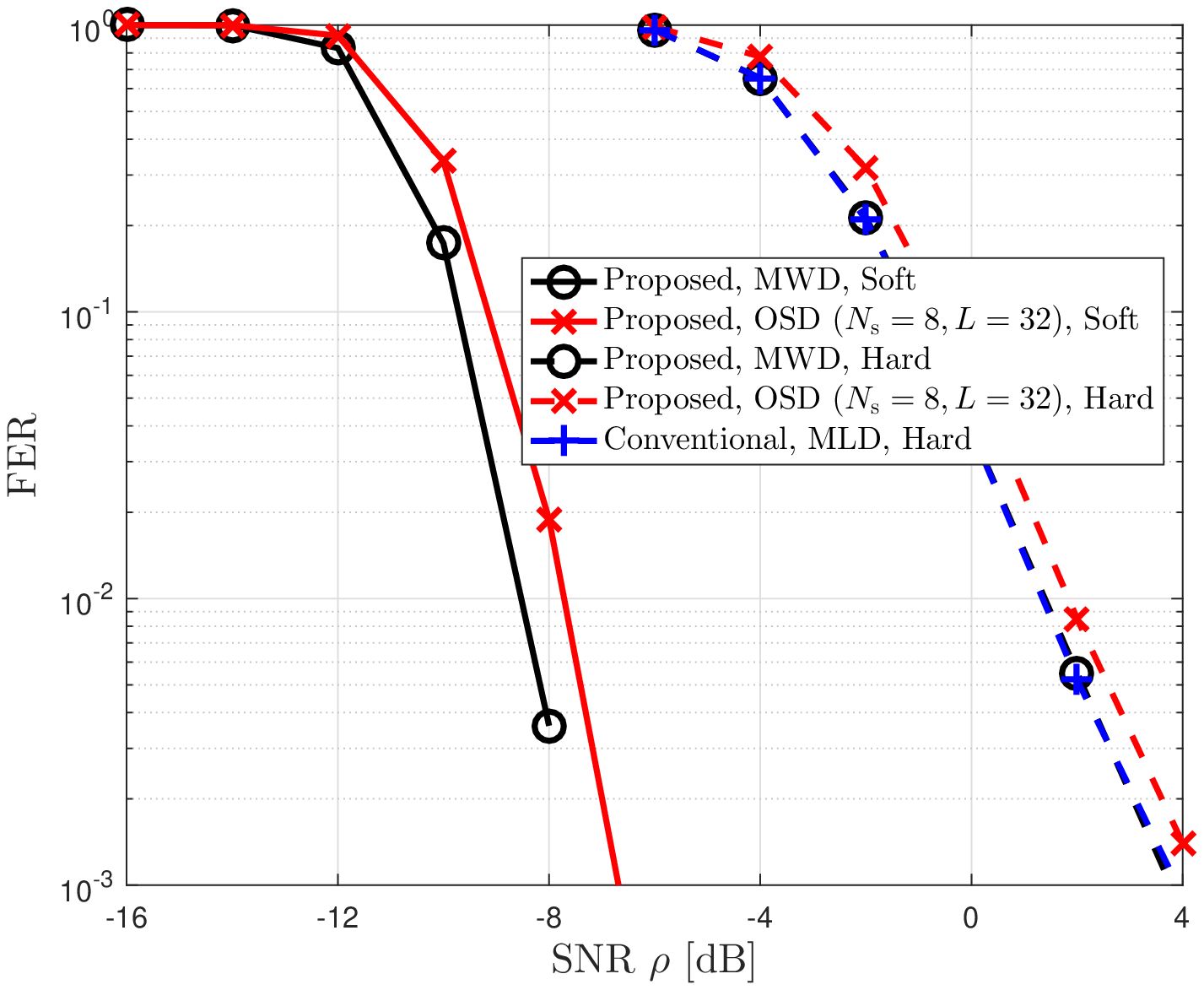} \vspace{-0.3cm}\caption{The FER vs. SNR of the proposed OSD, the proposed MWD, and the conventional MLD when either a soft detector or a hard detector is applied with $U=6$, $N=32$, 4-QAM modulation, and perfect CSIR.} \label{fig:Coded}
\end{figure}

Fig.~\ref{fig:Coded} compares the frame error rate (FER) of the OSD with those of the MLD and the MWD when CSIR is perfect. Fig.~\ref{fig:Coded} shows that for both the MWD and the OSD, almost 10-dB FER reduction is achieved by using the soft-input belief-propagation decoder instead of using the hard-input bit-flipping decoder. This result implies that the availability of the soft outputs has a significant impact on the detection performance when using the one-bit ADCs. In this context, the MWD and the OSD are suitable for the coded MIMO system with one-bit ADCs since soft outputs are available for both methods. Comparing two methods, the FER gap between them is less than 1 dB while the OSD reduces the detection complexity of the MWD by $72.6\%$ when $T_{\rm d}=N_{\rm B}$ (see Table~I). Therefore, the OSD provides a better performance-complexity tradeoff than the MWD also for the coded system.

\section{Conclusion}
In this paper we have proposed a new sphere decoding method for an uplink massive MIMO system with one-bit ADCs. One salient observation we found is that the weighted Hamming distance should be exploited to construct a list of codewords for sphere decoding due to the discrete nature of received signals.  We have also characterized the performance-complexity tradeoff achieved by the proposed OSD, in terms of its design parameters. Using simulations, we have shown that the proposed OSD effectively reduces the detection complexity of the MLD, while achieving near-MLD detection performance for both coded and uncoded systems.

An important direction for future research is to extend the proposed algorithm to frequency-selective channels in order to improve the practicality of the proposed algorithm. Another interesting extension is to develop a sphere decoding algorithm for an uplink massive MIMO system that uses low-resolution ADCs beyond one-bit precision. It would also be interesting to develop a low-complexity list construction method that may further reduce the complexity of the proposed algorithm.

\appendices

\section{Proof of Theorem~\ref{Thm:SEP}}\label{sec:Apdx1}
In this proof, we omit the index $t$ of time slot for ease of exposition. Suppose that the channel matrix of the system is given by ${\bf H}$. Then the weight vectors ${\bf w}_k$ and ${\bf \tilde w}_k$ of the proposed OSD are deterministic vectors from \eqref{eq:weight1} and \eqref{eq:weight2}, respectively. In this case, the probability that  the true symbol index does not belong to the codeword list in the sphere is expressed as
\begin{align}
\sum_{k=1}^K {\rm Pr}\left(k \notin \mathcal{S}({\bf y}), {\bf x} = {\bf x}_k\right)
& = \frac{1}{K}\sum_{k=1}^K {\rm Pr}\left( k \notin \mathcal{S}({\bf y}) \big| {\bf x} = {\bf x}_k \right), \label{eq:Apdx1:Definition}
\end{align}
provided that all possible symbol vectors are transmitted with an equal probability $\frac{1}{K}$.
By the definition of the codeword list in the sphere given in \eqref{eq:sphere_set}, the pair-wise probability in \eqref{eq:Apdx1:Definition} is
\begin{align}
{\rm Pr}\left( k \notin \mathcal{S}({\bf y}) | {\bf x} = {\bf x}_k \right)
&= {\rm Pr}\left( k \notin \bigcup_{g=1}^G \mathcal{S}_g\big({\bf y}^{(g)}\!,L\big) \vast| {\bf x} = {\bf x}_k \right) = \prod_{g=1}^G {\rm Pr}\left( k \notin \mathcal{S}_g\big({\bf y}^{(g)}\!,L\big) \Big| {\bf x} = {\bf x}_k \right), \label{eq:Apdx1:P_pair}
\end{align}
where the equality in \eqref{eq:Apdx1:P_pair} is obtained from the statistical independence of the noise vector in \eqref{eq:receive_real}.
An event $\{k \notin \mathcal{S}_g\big({\bf y}^{(g)},L\big)\}$ in \eqref{eq:Apdx1:P_pair} implies that the maximum weighted Hamming distance of the sub-codeword vector in $\mathcal{S}_g\big({\bf y}^{(g)},L\big)$ to ${\bf y}^{(g)}$ is less than that of the $k$-th sub-codeword vector. Using this fact, we rewrite the pair-wise probability in \eqref{eq:Apdx1:P_pair} as
\ifonecol
\begin{align}
&{\rm Pr}\left( k \notin \mathcal{S}_g\big({\bf y}^{(g)},L\big) \Big| {\bf x} = {\bf x}_k \right) \nonumber \\
&\leq {\rm Pr}\Bigg\{ \max_{j \in \mathcal{S}_{g}\big({\bf y}^{(g)},L\big)}	{d}_{\rm w}\left({\bf c}_{j}^{(g)},{\bf y}^{(g)};{\bf w}_{j}^{(g)}, {\bf \tilde w}_{j}^{(g)}\right) \leq {d}_{\rm w}\left({\bf c}_k^{(g)},{\bf y}^{(g)};{\bf w}_{k}^{(g)}, {\bf \tilde w}_{k}^{(g)}\right) \bigg| {\bf x} = {\bf x}_k \Bigg\},\label{eq:Apdx1:P_pair2}
\end{align}
\else
\begin{align}
&{\rm Pr}\left( k \notin \mathcal{S}_g\big({\bf y}^{(g)},L\big) \Big| {\bf x} = {\bf x}_k \right) \nonumber \\
&\leq {\rm Pr}\Bigg\{ \max_{j \in \mathcal{S}_{g}\big({\bf y}^{(g)},L\big)}	{d}_{\rm w}\left({\bf c}_{j}^{(g)},{\bf y}^{(g)};{\bf w}_{j}^{(g)}, {\bf \tilde w}_{j}^{(g)}\right) \nonumber \\
&~~~~~~~~~~~~\leq {d}_{\rm w}\left({\bf c}_k^{(g)},{\bf y}^{(g)};{\bf w}_{k}^{(g)}, {\bf \tilde w}_{k}^{(g)}\right) \bigg| {\bf x} = {\bf x}_k \Bigg\},\label{eq:Apdx1:P_pair2}
\end{align}
\fi
where the inequality in \eqref{eq:Apdx1:P_pair2} is due to the equality condition.

\ifonecol
\newcounter{tempeqn}
\begin{figure*}[t]
	\setcounter{tempeqn}{\value{equation}}
	\setcounter{equation}{41}
	{\small \begin{align} \label{eq:Apdx1:P_pair3}
		&{\rm Pr}\left( k \notin \mathcal{S}_g\big({\bf y}^{(g)},L\big) \Big| {\bf x} = {\bf x}_k \right) \nonumber \\
		&\leq \sum_{{\bf e}\in\{0,1\}^{N_{\rm s}}} {\rm Pr}\Bigg\{ \max_{j \in \mathcal{S}_{g}\big({\bf y}^{(g)}\!,L\big)}	{d}_{\rm w}\left({\bf c}_{j}^{(g)},{\bf y}^{(g)};{\bf w}_{j}^{(g)}, {\bf \tilde w}_{j}^{(g)}\right) \leq {d}_{\rm w}\left({\bf c}_k^{(g)},{\bf y}^{(g)};{\bf w}_{k}^{(g)}, {\bf \tilde w}_{k}^{(g)}\right) \bigg| {\bf x} = {\bf x}_k ,{\rm E}_k({\bf e})  \Bigg\} {\rm Pr}\left({\rm E}_k({\bf e})  \right) \nonumber \\
		&= \sum_{{\bf e}\in\{0,1\}^{N_{\rm s}}} {\mathbb I}\Bigg\{ \max_{j \in \mathcal{S}_{g} \big({\bf c}_k^{(g)}\!-2 {\bf e}\circ{\bf c}_k^{(g)}\!,L\big)}
		{d}_{\rm w}\left({\bf c}_{j}^{(g)},{\bf c}_k^{(g)};{\bf w}_{j}^{(g)}, {\bf \tilde w}_{j}^{(g)}\right)
		+ \sum_{i=1}^{N_{\rm s}} \left({w}_{j,i}^{(g)} - {\tilde w}_{j,i}^{(g)}\right)c_{k,i}c_{j,i}e_i \nonumber \\
		&\qquad\qquad\qquad\qquad\qquad\qquad\qquad\qquad\qquad\qquad\qquad\qquad \leq \sum_{i=1}^{N_{\rm s}} w_{k,i}^{(g)}e_{i} + {\tilde w}_{k,i}^{(g)}(1-e_i) \Bigg| {\bf x} = {\bf x}_k \Bigg\} {\rm Pr}\left({\rm E}_k({\bf e})  \right)
		\nonumber \\
		&=\sum_{{\bf e}\in\{0,1\}^{N_{\rm s}}} {\mathbb I}\vast\{ \max_{j \in \mathcal{S}_{g} \big({\bf c}_k^{(g)}\!-2 {\bf e}\circ{\bf c}_k^{(g)}\!,L\big)}
		{d}_{\rm w}\left({\bf c}_{j}^{(g)},{\bf c}_k^{(g)};{\bf w}_{j}^{(g)}, {\bf \tilde w}_{j}^{(g)}\right) + {\bf e}^\top {\bm \Delta}_{k,j}^{(g)}
		\leq  {\bf 1}_{N_{\rm s}}^\top {\bf \tilde w}_{k}^{(g)} \vast| {\bf x} = {\bf x}_k  \vast\} {\rm Pr}\left({\rm E}_k({\bf e}) \right).	
		\end{align}}
	\setcounter{equation}{\value{tempeqn}}
	\hrulefill	
\end{figure*}

We simplify \eqref{eq:Apdx1:P_pair2} by introducing the notion of an error vector ${\bf e}=[e_{1},e_{2},\cdots,e_{2N}]^\top\in\{0,1\}^{N_{\rm s}}$ where ${e}_i=1$ represents that the sign of the received signal is flipped due to the noise at the $i$-th position. Using the error vector, we denote ${\rm E}_k({\bf e})$ as an event that the received signal for the $g$-th sub-vector is given by
\begin{align}\label{eq:Apdx1:y_vec}
y_i^{(g)} =
\begin{cases}
-c_{k,i}^{(g)}, & e_i = 1, \\
c_{k,i}^{(g)}, & e_i = 0, \\
\end{cases}~\text{for}~i\in \{1,2,\ldots,N_{\rm s}\}.
\end{align}
Then we can rewrite \eqref{eq:Apdx1:P_pair2} as \eqref{eq:Apdx1:P_pair3} given at the top of this page,\setcounter{equation}{42}
\else
\newcounter{tempeqn}
\begin{figure*}[t]
	\setcounter{tempeqn}{\value{equation}}
	\setcounter{equation}{41}
	{\small \begin{align} \label{eq:Apdx1:P_pair3}
		&{\rm Pr}\left( k \notin \mathcal{S}_g\big({\bf y}^{(g)},L\big) \Big| {\bf x} = {\bf x}_k \right)
		\leq \sum_{{\bf e}\in\{0,1\}^{N_{\rm s}}} {\rm Pr}\left\{ \max_{j \in \mathcal{S}_{g}\big({\bf y}^{(g)}\!,L\big)}	{d}_{\rm w}\left({\bf c}_{j}^{(g)},{\bf y}^{(g)};{\bf w}_{j}^{(g)}, {\bf \tilde w}_{j}^{(g)}\right) \leq {d}_{\rm w}\left({\bf c}_k^{(g)},{\bf y}^{(g)};{\bf w}_{k}^{(g)}, {\bf \tilde w}_{k}^{(g)}\right) \bigg| {\bf x} = {\bf x}_k ,{\rm E}_k({\bf e})  \right\} {\rm Pr}\left({\rm E}_k({\bf e})  \right) \nonumber \\
		&= \sum_{{\bf e}\in\{0,1\}^{N_{\rm s}}} {\mathbb I}\left\{ \max_{j \in \mathcal{S}_{g} \big({\bf c}_k^{(g)}\!-2 {\bf e}\circ{\bf c}_k^{(g)}\!,L\big)}
		{d}_{\rm w}\left({\bf c}_{j}^{(g)},{\bf c}_k^{(g)};{\bf w}_{j}^{(g)}, {\bf \tilde w}_{j}^{(g)}\right)
		+ \sum_{i=1}^{N_{\rm s}} \left({w}_{j,i}^{(g)} - {\tilde w}_{j,i}^{(g)}\right)c_{k,i}c_{j,i}e_i
		\leq \sum_{i=1}^{N_{\rm s}} w_{k,i}^{(g)}e_{i} + {\tilde w}_{k,i}^{(g)}(1-e_i) \Bigg| {\bf x} = {\bf x}_k \right\} {\rm Pr}\left({\rm E}_k({\bf e})  \right)
		\nonumber \\
		&=\sum_{{\bf e}\in\{0,1\}^{N_{\rm s}}} {\mathbb I}\vast\{ \max_{j \in \mathcal{S}_{g} \big({\bf c}_k^{(g)}\!-2 {\bf e}\circ{\bf c}_k^{(g)}\!,L\big)}
		{d}_{\rm w}\left({\bf c}_{j}^{(g)},{\bf c}_k^{(g)};{\bf w}_{j}^{(g)}, {\bf \tilde w}_{j}^{(g)}\right) + {\bf e}^\top {\bm \Delta}_{k,j}^{(g)}
		\leq  {\bf 1}_{N_{\rm s}}^\top {\bf \tilde w}_{k}^{(g)} \vast| {\bf x} = {\bf x}_k  \vast\} {\rm Pr}\left({\rm E}_k({\bf e}) \right).		 
		\end{align}}
	\setcounter{equation}{\value{tempeqn}}
	\hrulefill	
\end{figure*}

We simplify \eqref{eq:Apdx1:P_pair2} by introducing the notion of an error vector ${\bf e}=[e_{1},e_{2},\cdots,e_{2N}]^\top\in\{0,1\}^{N_{\rm s}}$ where ${e}_i=1$ represents that the sign of the received signal is flipped due to the noise at the $i$-th position. Using the error vector, we denote ${\rm E}_k({\bf e})$ as an event that the received signal for the $g$-th sub-vector is given by
\begin{align}\label{eq:Apdx1:y_vec}
y_i^{(g)} =
\begin{cases}
-c_{k,i}^{(g)}, & e_i = 1, \\
c_{k,i}^{(g)}, & e_i = 0, \\
\end{cases}~\text{for}~i\in \{1,2,\ldots,N_{\rm s}\}.
\end{align}
Then we can rewrite \eqref{eq:Apdx1:P_pair2} as \eqref{eq:Apdx1:P_pair3} given at the top of the next page,\setcounter{equation}{42}
\fi
where ${\bm \Delta}_{k,j}^{(g)} \in \mathbb{R}^{N_{\rm s}}$ is a vector whose $i$-th element is
\begin{align}\label{eq:Apdx1:Delta}
{\Delta}_{k,j,i}^{(g)} =  \left({w}_{j,i}^{(g)} - {\tilde w}_{j,i}^{(g)}\right)c_{k,i}c_{j,i}  -  \left({w}_{k,i}^{(g)} - {\tilde w}_{k,i}^{(g)}\right).
\end{align}
Let $\mathcal{\bar E}_{k}^{(g)}(L)$ be a set of all vectors in $\{0,1\}^{N_{\rm s}}$ that satisfy the inequality condition of the indicator function in \eqref{eq:Apdx1:P_pair3}, i.e.,
\begin{align}\label{eq:Apdx1:E_k}
\mathcal{\bar E}_{k}^{(g)}(L) = \left\{{\bf e} :\max_{j \in \mathcal{S}_{g} \big({\bf c}_k^{(g)}\!-2 {\bf e}\circ{\bf c}_k^{(g)}\!,L\big)} ~d_{k,j}^{(g)} + {\bf e}^\top {\bm \Delta}_{k,j}^{(g)} , ~{\bf e} \in \{0,1\}^{N_{\rm s}}   \right\},
\end{align}
where $d_{k,j}^{(g)}= {d}_{\rm w}\left({\bf c}_{j}^{(g)},{\bf c}_k^{(g)};{\bf w}_{j}^{(g)}, {\bf \tilde w}_{j}^{(g)}\right)$ for all $j,k,g$.
Using this set, \eqref{eq:Apdx1:P_pair3} is expressed in a simplified form:
\begin{align} \label{eq:Apdx1:P_pair4}
{\rm Pr}\left( k \notin \mathcal{S}_g\big({\bf y}^{(g)}\!,L\big) \Big| {\bf x} = {\bf x}_k \right)
\leq \sum_{{\bf e}\in\mathcal{\bar E}_{k}^{(g)}(L) } {\rm Pr}\left({\rm E}_k({\bf e}) \right).	
\end{align}
We derive the upper bound of \eqref{eq:Apdx1:P_pair4} by constructing an extended set $\mathcal{E}_{k}^{(g)}(L)$ which contains the set $\mathcal{\bar E}_{k}^{(g)}(L)$ in \eqref{eq:Apdx1:P_pair4} as a subset, i.e., $\mathcal{\bar E}_{k}^{(g)}(L) \subset \mathcal{E}_k^{(g)}(L) $. For this, let $d_{{\rm min},k}^{(g)}({\bf e},L)$ be the $L$-th smallest element of a set $\Big\{d_{k,j}^{(g)} + {\bf e}^\top {\bm \Delta}_{k,j}^{(g)},~j \in \mathcal{K}\setminus \{k \}\Big\}$.
Then because
\begin{align}
d_{{\rm min},k}^{(g)}({\bf e},L) \leq \max_{j \in \mathcal{S}_{g} \big({\bf c}_k^{(g)}\!-2 {\bf e}\circ{\bf c}_k^{(g)}\!,L\big)} ~d_{k,j}^{(g)} + {\bf e}^\top {\bm \Delta}_{k,j}^{(g)},
\end{align}
we can construct the extended set $\mathcal{E}_{k}^{(g)}(L)$ of $\mathcal{\bar E}_{k}^{(g)}(L)$ as
\begin{align}\label{eq:Apdx1:E_k_prime}
\mathcal{E}_{k}^{(g)}(L) = \bigg\{{\bf e} : &~ d_{{\rm min},k}^{(g)}({\bf e},L) \leq {\bf 1}_{N_{\rm s}}^\top {\bf \tilde w}_{k}^{(g)}, {\bf e} \in \{0,1\}^{N_{\rm s}}   \bigg\}.
\end{align}
Using this extended set, we rewrite \eqref{eq:Apdx1:P_pair4} as
\begin{align} \label{eq:Apdx1:P_pair_upp}
{\rm Pr}\left( k \notin \mathcal{S}_g\big({\bf y}^{(g)}\!,L\big) \Big| {\bf x} = {\bf x}_k \right)
\leq \sum_{{\bf e}\in\mathcal{E}_{k}^{(g)}(L) } {\rm Pr}\left({\rm E}_k({\bf e}) \right).	
\end{align}

Now, the remaining term in \eqref{eq:Apdx1:P_pair_upp} is the probability of the event ${\rm E}_k({\bf e})$. By defining a set $\mathcal{I}({\bf e}) = \{i: e_i=1\}$, this probability can be represented as
\begin{align}
{\rm Pr}\left({\rm E}_k({\bf e})  \right) &= \!\prod_{i\in \mathcal{I}({\bf e})} \!\!{\rm Pr}(c_{k,i}\neq y_i | {\bf x}={\bf x}_k) \!\! \prod_{i\notin\mathcal{I}({\bf e})} \!\!{\rm Pr}(c_{k,i}= y_i | {\bf x}={\bf x}_k)\nonumber \\
&= \!\prod_{i\in \mathcal{I}({\bf e})} \! Q\left(\! \sqrt{\frac{2}{\sigma^2}}|{\bf h}_i^\top {\bf x}_k|  \right)
\!\prod_{i\notin \mathcal{I}({\bf e})} \! \left\{ 1- Q\left(\! \sqrt{\frac{2}{\sigma^2}}|{\bf h}_i^\top {\bf x}_k|  \right) \right\}.  \label{eq:Thm2:Pr_Ek}
\end{align}
From Lemma~1, we can approximate the right-hand-side of \eqref{eq:Thm2:Pr_Ek} using two weights $w_{k,i}$ and  ${\tilde w}_{k,i}$ as follows:
\begin{align}\label{eq:Apdx1:Pr_Ek_upp}
{\rm Pr}\left({\rm E}_k({\bf e})  \right)
&\approx \exp\left(-\sum_{i\in \mathcal{I}({\bf e})}w_{k,i} - \sum_{i\notin \mathcal{I}({\bf e})}{\tilde w}_{k,i} \right) = \exp\left( -{\bf e}^\top{\bf w}_k^{(g)} - ({\bf 1}_{N_{\rm s}}-{\bf e})^\top{\bf \tilde w}_k^{(g)}  \right).
\end{align}
It is noticeable that the approximation used in \eqref{eq:Apdx1:Pr_Ek_upp} is tight because the Q-function approximation in Lemma~1 has a bounded error less than $10^{-3}$.
Plugging \eqref{eq:Apdx1:Pr_Ek_upp} into \eqref{eq:Apdx1:P_pair_upp} and then applying the result to \eqref{eq:Apdx1:P_pair} and \eqref{eq:Apdx1:Definition} yields the approximate upper bound in \eqref{eq:Thm2}; this completes the proof.


\end{document}